\RequirePackage{fix-cm}
\documentclass[smallcondensed]{svjour3}     
\smartqed  
\usepackage{graphicx}
\usepackage{times}
\usepackage{mathptmx}      
\usepackage{amsmath}
\usepackage{amsfonts}
\usepackage{amssymb}
\usepackage{bbm}

\usepackage[ruled,noend,lined]{algorithm2e}

\usepackage[final]{pdfpages}

\usepackage{booktabs}
\usepackage{rotating}
\usepackage{multirow}
\usepackage{array}

\usepackage{graphicx}
\usepackage{subfig}
\usepackage{epstopdf}
\DeclareGraphicsExtensions{.pdf,.eps,.png,.jpg,.tif,.tiff,.ps}
\graphicspath{{img/}, {img/new-graphics/}, {.}}

\usepackage{url}
\usepackage{hyperref}
\usepackage{xspace}

\newcommand{\KMeans}{\mbox{K-Means}}%
\newcommand{\TDCKMeans}{\mbox{TDCK-Means}}%
\newcommand{\cluspath}{\mbox{ClusPath}}%

\definecolor{navy}{rgb}{0.1, 0.1, 0.8}
\definecolor{gray}{rgb}{0.4, 0.4, 0.4}

\newcommand{\eat}[1]{}
\newcommand{\rev}[1]{{\color{black}{#1}}}

\newenvironment{aligntextstyle*}{%
  \let\temp\displaystyle
  \let\displaystyle\textstyle
  \start@align\@ne\st@rredtrue\m@ne
}{%
  \endalign
  \let\displaystyle\temp
}

\newcommand{\squishlist}{
 \begin{list}{$\bullet$}
  { \setlength{\itemsep}{0pt}
     \setlength{\parsep}{3pt}
     \setlength{\topsep}{3pt}
     \setlength{\partopsep}{0pt}
     \setlength{\leftmargin}{1.5em}
     \setlength{\labelwidth}{1em}
     \setlength{\labelsep}{0.5em} } }

\newcommand{\squishlisttwo}{
 \begin{list}{$\bullet$}
  { \setlength{\itemsep}{0pt}
    \setlength{\parsep}{0pt}
    \setlength{\topsep}{0pt}
    \setlength{\partopsep}{0pt}
    \setlength{\leftmargin}{1.5em}
    \setlength{\labelwidth}{1.5em}
    \setlength{\labelsep}{0.5em} } }

\newcommand{\squishend}{
  \end{list}  }

\hypersetup
{
pdftitle={A Temporal-driven Clustering Solution to Inferring Typical Evolution Paths},%
pdfauthor={Marian-Andrei Rizoiu, Julien Velcin, St\'phane Bonnevay and St\'ephane Lallich},%
pdfkeywords={semi-supervised clustering, temporal clustering, temporal cluster graph structure, evolutionary optimization, Pareto front estimation.}%
}

\journalname{Data Mining and Knowledge Discovery}
\begin{document}

\title{\cluspath{}: A Temporal-driven Clustering to Infer Typical Evolution Paths}

\titlerunning{Inferring Typical Evolution Paths}  

\author{Marian-Andrei Rizoiu \and Julien Velcin \and \\
St\'ephane Bonnevay \and St\'ephane Lallich}
\authorrunning{Marian-Andrei Rizoiu et al.} 

\institute{Marian-Andrei Rizoiu \at
              NICTA \& Australian National University, 7 London Circuit, Canberra, Australia.\\
              \email{Marian-Andrei.Rizoiu@nicta.com.au}       
           \and
           Julien Velcin \and St\'ephane Bonnevay \and St\'ephane Lallich \at
              ERIC laboratory, Universit\'e de Lyon, Lyon, France.\\
			  \email{Julien.Velcin@univ-lyon2.fr, Stephane.Bonnevay@univ-lyon1.fr,\\ Stephane.Lallich@univ-lyon2.fr}
}

\date{Received: date / Accepted: date}

\maketitle

\begin{abstract}
We propose \cluspath{}, a novel algorithm for detecting general evolution tendencies in a population of \rev{entities}.
We show how abstract notions, such as the Swedish socio-economical model (in a political dataset) or the companies fiscal optimization (in an economical dataset) can be inferred from low-level descriptive features.
Such high-level regularities in the evolution of entities are detected by combining spatial and temporal features into a spatio-temporal dissimilarity measure and \textcolor{black}{using} semi-supervised clustering techniques.
The relations between the evolution phases are modeled using a graph structure, inferred simultaneously with the partition, by using a ``slow changing world'' assumption.
The idea is to ensure a smooth passage for entities along their evolution paths, which catches the long-term trends in the dataset.
\textcolor{black}{Additionally,} we also provide a method, based on an evolutionary algorithm, to tune \rev{the parameters of \cluspath{}} to new, unseen datasets.
This method assesses the fitness of a solution using four opposed quality measures and proposes a balanced compromise.
%
%

\end{abstract}

\vspace*{-0.02\textheight}
\section{Introduction}

\setlength{\textfloatsep}{1\baselineskip}
\setlength{\floatsep}{1\baselineskip}
\setlength{\intextsep}{1\baselineskip}

Knowledge is often hidden in plain view, within the sheer amount of available data.
Refining data into information by discovering patterns is \eat{one of }the main purposes of Data Mining.
In this paper, we are interested in the more specific problem of discovering general temporal trends, also known as typical evolution paths.
This makes the problem of pattern mining more difficult, by adding to it the temporal dimension of data.
It changes the definition of the learning problem, since the description of entities is temporally contextualized.
We study a population of entities, described over a period of time by low-level descriptive features.
Our final aim is to detect the typical paths of evolution taken by most of these entities over the extent of recorded time.
Considering that an entity's description can change over time, we define an \textit{evolution phase} as a period of limited extent in time during which multiple \rev{entities} in the dataset share similar descriptions.
Therefore, to be informative, evolution phases should be coherent in time and in the descriptive space.
An \textit{evolution path} is defined as a succession of evolution phases followed by a large number of entities in the dataset, which can also be seen as a typical trajectory through the evolution space.
Such general evolution trends can reveal information about the more complex hidden phenomena which happen in the population of entities.
We show, in Sect.~\ref{sec:experiments}, that high-level concepts such as socio-political regimes or fiscal strategies can be detected from the low-level descriptive features.
For example, the \rev{``Swedish Social and Economical Model''~\cite{Erixon2000}} can be mapped on the evolution path followed by the northern European countries, in a dataset described using features such as debt-to-GDP ratio, unemployment rate and political coloring of the parliament.
Similarly, we detect a strong trend in a population of companies, in which significantly less tax is payed, while the net income increases: the fiscal optimization of international corporations.

To this end, we propose \cluspath{}, an algorithm that organizes a set of observations into a structured partition, coherent both in the descriptive space and in the temporal space.
We use a temporal-aware dissimilarity measure for assessing the similarity between observations.
Furthermore, we use a semi-supervised technique using must-link pairwise constraints to ensure the contiguous segmentation of observations associated to an entity.
The resulted clusters are interpreted as evolution phases.
\textcolor{black}{
The main novelty of \cluspath{} is that the three components of the clusters: i) descriptive, ii) temporal and iii) the graph of relations between them are inferred simultaneously, in one optimization procedure.
This creates an intertwining, which allows the three components to influence each other, during the optimization process.
The major advantage over other approaches (like co-clustering or post-clustering structure inference) is that it allows more flexibility during the optimization process and creates a partition more adequate to describe the data.
Furthermore,
} it ensures that the entities have a smooth passage between the phases on the evolution path.
While we run the risk of losing non-smooth \rev{entity} evolutions, the purpose of our application is to detect the evolution paths followed by the majority of the population.
This is due to the ``slow changing world'' assumption, which states that the long-term trends detectable in a population have a higher inertia and they evolve more slowly than \rev{entities}.
This assumption might not be desirable for all contexts, for example i) in applications in which it is important to capture the fined-grained evolutions of \rev{entities}. (\textit{e.g.}, stock exchange market) or ii) in which the general trend does not possess a high inertia (\textit{e.g.}, popularity in social networks, online memes \textit{etc.}).
For these applications, \cluspath{} \rev{has a parameter (\textit{i.e.} $\lambda_2$, defined in Sect.~\ref{subsec:construct-objective-measure}) which} allows the modulation of the degree in which this hypothesis is enforced.
Additionally, we propose \textcolor{black}{an optional} method, based on an evolutionary algorithm, to automatically tune \rev{the parameters of \cluspath{}} on new, unseen datasets.

The remainder of this paper is structured as follows. 
In Sect.~\ref{sec:state-of-the-art}, we present some previous related work. 
In Sect.~\ref{sec:our-proposal}, we define the learning objectives, we translate them into an optimization problem, we introduce \cluspath{} and the evolutionary heuristic for automatically tuning the values of parameters.
Sect.~\ref{sec:experiments} presents the datasets we use, the quality measures and the performed experimentations.
We conclude in Sect.~\ref{sec:conclusion} and plan some future work.

\vspace*{-0.03\textheight}
\section{State of the Art}
\label{sec:state-of-the-art}
\vspace*{-0.01\textheight}

The purpose of \textit{Evolutionary clustering} is to capture the temporal evolution of clusters, given data observations and their creation time.
A good clustering result should fit the current data well, while simultaneously not deviate too dramatically from the recent history~\cite{CHI07}.
Initial frameworks have been designed for distance-based clustering, such as \KMeans{} and agglomerative hierarchical clustering~\cite{CHA06a}.
\cite{CHI07} have extended the evolutionary framework to spectral clustering with the emphasis on smoothing clustering results over time to avoid sudden changes.
\cite{XU12} take a generative models approach and apply it to dynamic social network analysis.
All of these evolutionary clustering approaches rely on time discretization into temporal windows of arbitrary size, whereas \cluspath{} integrates the temporal dimension as a variable, without the need of discretization.

\TDCKMeans{}~\cite{RIZ12} was introduced to detect typical evolution phases in a population of entities.
The main contributions of this work were i) proposing a temporal-aware dissimilarity measure, used to assess the similarity between two observations, both in the multidimensional descriptive and temporal spaces and ii) assuring a contiguous segmentation by imposing semi-supervised must-link constraints~\cite{WAG01} and a continuous time-dependent penalty function for breaking the constraints.
%
Other algorithms in the literature use constraints for segmentation purposes.
tcK-Means~\cite{LIN06} uses
must-link constraints and inflicts a fixed penalty when the following conditions are fulfilled simultaneously: the observations are not assigned to the same cluster and the time difference between their timestamps is less than some threshold.
Similarly, \rev{\cite{TOR07}
detect tasks} performed during a day, based on video, sound and GPS data.
Aligned Cluster Analysis~\cite{Araujo2014} is an extension of the
kernel k-means clustering algorithm, in which side information is added in the form of pairwise constraints to its objective function.
Its purpose is segmenting time-series and \rev{clustering} them together.

A recent extension of \TDCKMeans{}~\cite{Rizoiu2014b} proposes an \textit{a posteriori} method for organizing the constructed clusters as a graph.
The construction is based on the transitions of entities between phases.
While the aim of \cluspath{} is also to identify the links between clusters, it differs fundamentally from the \textit{a posteriori} construction by inferring the relations \emph{simultaneously \rev{while} clustering} of the observations.
This creates an intertwining, by allowing the partitioning to influence the structure of clusters, and, conversely, the links between clusters influence the assignment of observations to clusters.

\cluspath{} infers the relations between clusters by combining multiple criteria into an objective function and optimizing it using a gradient descent method.
A related learning problem is \textit{relational multicriteria clustering}.
The aim is to detect clusters \rev{of alternatives} in a multicriteria context and to identify relations between these clusters. 
In~\cite{ROC13}, a classical clustering is first applied \rev{to the set of possible alternative and each cluster is evaluated using predefined measures.
A partial order outranking relation is established between clusters, based on the scores of the evaluation measures and the preferences of the decision maker.
The outranking relations are constructed as a \emph{post-treatment} (after clustering), using a multi-criteria pair-wise comparison procedure.}
\cite{SME09} use a distance measure that is based on binary preference relations between different alternatives.
The distance is extended to construct a binary outranking matrix between clusters.
The outranking matrix is constructed at each clustering iteration, but it has no influence over the assignment of actions to clusters and it is calculated solely based on the composition of clusters.
%

A distinct, but somewhat related field is that of clustering of multi-dimensional trajectories.
The crucial difference between this field and our work is that the former usually seeks to find similarities between entire multi-dimensional data series (\textit{e.g.}, storm path trajectories, drug therapy response) in order to find connections between the evolution of different entities.
These approaches usually treat the entire temporal series as single data points.
For example, \cite{Gaffney1999} model the set of trajectories as individual sequences of points generated from a finite mixture model.
\cite{Liang2013} predict glaucoma evolution in patients by using previous recorded disease evolutions.
\rev{
In the first step, clustering is applied to gather patients similar to the target patient. 
The second step fits a predictive model on the set of patients found in the first step, and predicts the future disease condition}.
In \cite{Siddiqui2012}, a mixture model of Markov Chains is learned and used to predict the next most likely state/cluster per object.
Apart from serving a different purpose, the first approach is unsupervised, the second and the third are supervised, whereas \cluspath{} is a semi-supervised algorithm.
\cite{Kalnis2005} approach a related learning problem: detecting of trajectories of moving clusters.
Their underlying assumption \rev{is that the data contains} dense groups of individuals which move together in space and time (\textit{i.e.}, the moving clusters).
They construct individual partitions at each timestep and detecting pairs of clusters in successive timesteps, susceptible of belonging to the same moving cluster.
Our problem differs mainly because an evolution path is not a unitary entity as a moving cluster.
The individual evolution phases have arbitrary extents of time and each one can be part of multiple evolution phases.
Unlike the temporal instantiations of a single moving cluster, evolution phases have meaning by themselves and they are loosely connected in evolution paths.

\section{Our Proposal}
\label{sec:our-proposal}
\vspace*{-0.01\textheight}

\subsection{Formalization and learning objectives}
\label{subsec:formalization}
\vspace*{-0.01\textheight}

\rev{\textbf{Definitions and intuitions.}}
Each studied entity $ \phi_l \in \Phi$ is described using multiple attributes, which form the multidimensional description space.
To each entity correspond $N$ observations $(entity,$ $timestamp,$ $description)$.
An observation $x_i = (\phi_l, t_m, x_i^d), i \in 1..N$ means that the entity $\phi_l$ is described by the vector $x_i^d$ at the moment of time $t_m$.
To identify the entity associated with a particular observation $x_i$, we use the notation $x_i^\phi$.
Therefore the notations $x_i^\phi$ and $\phi_l$ both denote the entities, and we use one or the other depending on the point of view (\textit{i.e.}, observation- or entity- oriented).
Similarly, $x_i^t$ is the timestamp associated with the observation $x_i$.
Our learning problem, starting from such a dataset, aims at detecting typical evolution \textcolor{black}{phases and evolution} paths. 
There is a double interest: 
a) obtaining a broader understanding of the phases that the collection of entities went through over time (\textit{e.g.}, detecting the periods of global political instability, economic crisis, wealthiness \textit{etc.});
b) constructing the trajectory of an entity through the different phases (\textit{e.g.} a country may have gone through a period of military dictatorship, followed by a period of wealthy democracy).
We define an evolution phase $\mathcal{C}$ as a set of observations $x_i$, \rev{so that observations belonging to $\mathcal{C}$ are as similar (in terms of a similarity function) as possible among themselves and dissimilar to observations in other phases.
Unlike in classical clustering (for example \KMeans{}~\cite{MAC67}), observations should be similar both in the descriptive and in the temporal space.}
We consider that each entity $\phi_l$ is associated at every moment of time with one and only one evolution phase, \textit{i.e.}, each observation $x_i$ belongs to a single evolution phase.
Furthermore, phases are assumed to be linked to each other, to allow entities to temporally navigate between them.
A temporal succession of evolution phases forms an \emph{evolution path}.
Therefore, the ``slow changing world'' hypothesis, which assumes that long-term trends evolve slowly, translates into evolution paths in which successive evolution phases have high connection strengths.

\rev{
\textbf{Prototypes.}
We define $\mu = (\mu^t, \mu^d), \mu \in \mathcal{M}$ the prototype of an evolution phase $\mathcal{C}$, where $\mathcal{M}$ is the set of all prototypes.
Just like centroids in traditional clustering, prototypes behave likes ``central tendencies'' of their phases and characterize the phases both in the temporal and in the descriptive space.
Unlike centroids, the prototypes cannot be rigorously defined outside the learning problem, as they are dependent not only on the observations assigned with a phase, but also on the other prototypes and the choice of parameters (shown in Sect.~\ref{subsec:cluspath-algo}).
}

\begin{figure}[tbp]
	\centering
	\subfloat[] {
		\includegraphics[width=0.465\textwidth]{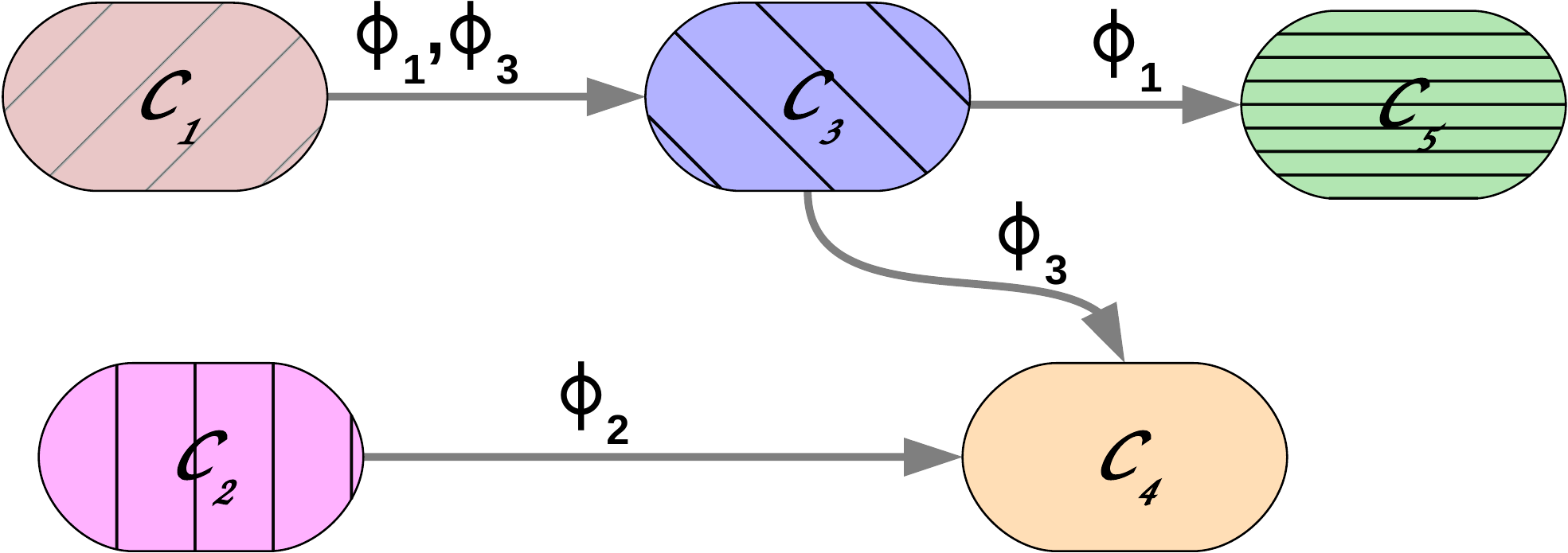}
		\label{subfig:cluster-structuring}
	}
	\quad
	\subfloat[] {
		\includegraphics[width=0.465\textwidth]{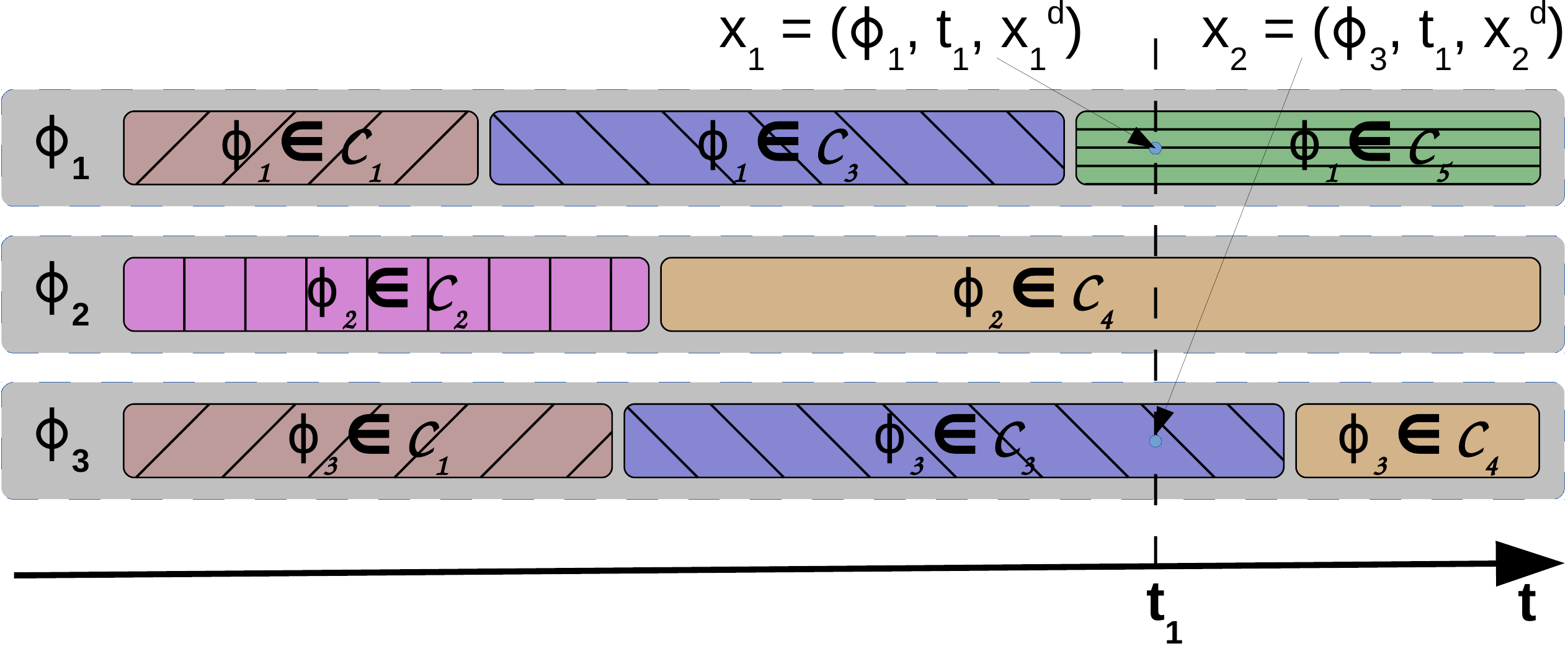}
		\label{subfig:entity-contiguity}
	}
	\caption{
	\rev{An example of a desired output, in which the evolutions of 3 entities ($\phi_1, \phi_2$ and $\phi_3$) are described using 5 phases ($\mathcal{C}_i, i = 1,\dots,5$). 
	For example, the evolution path of $\phi_3$ is $\mathcal{C}_1 \rightarrow \mathcal{C}_3 \rightarrow \mathcal{C}_3$.
	(a) The graph structure of the evolution phases.
	The arcs between two phases $(\mathcal{C}_i, \mathcal{C}_j)$ are labeled with the entities presenting the transition $\mathcal{C}_i \rightarrow \mathcal{C}_j$.
	(b) The observations of the 3 entities are partitioned contiguously into the 5 phases.}
	}
	\label{fig:desired-output}
\end{figure}

\rev{\textbf{Method.}
We infer typical evolution paths, by clustering the observations corresponding to entities into $k$ clusters, which serve as \emph{evolution phases}.
}
The links between the evolution phases are represented using a graph structure, defined by its adjacency matrix $ A = (a_{ij}) $, where $a_{i,j} \in [0, 1]$ is the strength of the link between clusters $\mathcal{C}_i$ and $\mathcal{C}_j$.
A value of $0$ denotes the absence of a link.
\rev{The connection strength of the link is proportional to i) the similarity of their prototypes $\mu_i$ and $\mu_j$, both in the temporal and descriptive space, and ii) the number of entities presenting the passage from  $\mathcal{C}_i$ to $\mathcal{C}_j$}.
The graph is oriented and, therefore, the matrix $A$ is not symmetrical.
Fig.~\ref{fig:desired-output} shows the desired result of our clustering algorithm.
Fig.~\ref{subfig:cluster-structuring} shows how the \rev{phases} $\mathcal{C}_j$ are structured into a graph structure and, in Fig.~\ref{subfig:entity-contiguity}, the series of observations belonging to each entity are assigned to \rev{phases}, thus forming continuous segments.
The succession of segments is interpreted as the entity's evolution path. 

We define the following objectives for the resulting partition:
\squishlisttwo    
	\item \rev{\textbf{\textit{Obj. 1}}: construct clusters which are coherent in the temporal and the descriptive space.
	Observations under a cluster should have similar descriptions (just as traditional clustering does) and they should be temporally close.
	Each cluster should provide a trade-off between the temporal and descriptive coherence, since the two might be contradictory.
	For example, two different periods with similar evolutions (\textit{e.g.}, two economical crises) should be regrouped separately, as they represent two distinct evolution phases;}
	\item \rev{\textbf{\textit{Obj. 2}}:} segment, as contiguously as possible, the series of observations for each entity. 
	The sequence of segments is interpreted as the entity's evolution path;
	\item \rev{\textbf{\textit{Obj. 3}}:} present smooth passages between phases on evolution paths.
	An evolution path should take an entity through highly similar evolution phases, \textit{i.e.}, changes should come in small increments.
\squishend     

\vspace*{-0.03\textheight}
\subsection{Constructing the Objective Measure}
\label{subsec:construct-objective-measure}
\vspace*{-0.01\textheight}

\cluspath{} fulfills the aforementioned objectives by a) translating them into several objectives and combining them into an overall objective function $\mathcal{J}$ and b) applying a gradient descent optimization method, in which $\mathcal{J}$ is minimized, using a \KMeans{}-like iterative relocation framework.
To optimize \textit{\textbf{Obj 1}}\eat{and \textbf{\textit{Obj 2}}}, we use the temporal-aware dissimilarity measure proposed by~\cite{RIZ12}:
\begin{equation} \label{eq:tuned-temp-distance}
	|| x_i - x_j||_{TA} = 1 - \left(1 - \gamma_d \frac{||x_i^d - x_j^d||^2}{\Delta d_{max}^2}\right)\left(1 - \gamma_t \frac{||x_i^t - x_j^t||^2}{\Delta t_{max}^{2}}\right)  
\end{equation}
where $\gamma_d$ and $\gamma_t$ are controlled by the parameter $\alpha \in [-1, 1]$, which acts like a slider, favoring the temporal component for $\alpha = -1$ or the descriptive component for $\alpha = 1$:
\begin{equation} \label{eq:gamma-d-gamma-t}
    \gamma_d= 
		\begin{cases}
   			1 + \alpha ,& \text{if } \alpha \in [-1, 0] \\
    		1, 			& \text{if } \alpha \in (0, 1]
    	\end{cases} 
	\;\;;\;\;
	\gamma_t= 
		\begin{cases}
			1 ,		   & \text{if } \alpha \in [-1, 0] \\
			1 - \alpha,& \text{if } \alpha \in (0, 1]
	\end{cases}  \enspace,
\end{equation}
$||\bullet||_{TA} \in [0,1]$ and a value of zero means identical observations.
$\Delta d_{max}$ and $\Delta t_{max}$ are the diameters of the descriptive and temporal space respectively (the largest distance encountered between two observations).
The temporal-aware measure allows to simultaneously take into account the similarity in the descriptive and temporal spaces, between two observations.
By using this measure, \cluspath{} minimizes the term:
$
	\sum_{\mu_p \in \mathcal{M}}\sum_{x_i \in \mathcal{C}_p} || x_i - \mu_p||_{TA} \enspace,
$
where $\mu_p$ is the abstraction of cluster $\mathcal{C}_p$, to which the observation $x_i$ is assigned.

%
The \rev{second objective (\textbf{\textit{Obj. 2}})} states that, for comprehension reasons, the series of observations belonging to an entity should be segmented contiguously.
We extend the initial segmentation mechanism of \TDCKMeans{}, by setting \textit{temporally-oriented must-link soft constraints} 
\textcolor{black}{
between each pair of observations, belonging to \rev{the same entity}. 
\rev{Each entity is associated with $N$ observations, therefore} each observation is involved in $N-1$ constraints, linking it to all the other observations. 
Each constraint $x_i$ has a direction, \textit{i.e.}, it is temporally-oriented.
$i-1$ of these constraints are incoming (from preceding observations to $x_i$), while $N-i$ are outgoing, towards the subsequent observations. 
A total of $(N-1)(N-2)/2$ constraints are set for each \rev{entity}.
A must-link constraint indicates that the two observations should be placed into the same cluster. 
Being soft constraints, \cluspath{} is allowed to break any number of them, while a time-decaying penalty is inflicted for each violation.
The penalty is more severe when the observations are closer in time and less severe when the two assigned clusters have a strong link (high value for $a_{i,j}$).
}
The penalty function is:
\begin{equation}\label{eq:penalty-function}
	w(x_i, x_k) = \beta * e^{-\frac{1}{2} \left( \frac{||x_i^t - x_k^t||}{\delta} \right)^2} \left( 1- a_{j,l}^2 \right) \mathbbm{1} \left[x_i^\phi = x_k^\phi \right] \mathbbm{1} \left[x_i^t < x_k^t \right] \enspace, 
\end{equation}
where $x_i \in \mathcal{C}_j$, $x_k \in \mathcal{C}_l$ and $\beta \in \mathbb{R^+}$ is the weight of the penalty function;
$\delta \in \mathbb{R^+}$ is a parameter which controls the width of the function.
The penalty function in Eq.~\ref{eq:penalty-function} is inspired from the Normal Distribution function and it does not require the discretization of time.
Respecting all constraints involves associating all observations of an \rev{entity} to the same evolution phase.
Therefore, the segmentation mechanism strives to acquire a trade-off between clustering observations together and putting them into separate, yet well connected clusters.
We obtain the first term ($T_1$) of the objective function, dealing with assigning observations to clusters:
\begin{equation} \label{eq:term_1}
	T_{1} =  \sum_{\mu_p \in \mathcal{M}}\sum_{x_i \in \mathcal{C}_p} \left( || x_i - \mu_p||_{TA} + \sum_{\substack{x_k \in \mathcal{C}_q \\ q \neq p , x_i^\phi = x_k^\phi}}^{x_i^t < x_k^t} \beta * e^{-\frac{1}{2} \left( \frac{||x_i^t - x_k^t||}{\delta} \right)^2} \left(1-a_{p,q}^2 \right) \right) \enspace.
\end{equation}

The following terms of the objective function, leverage the influence of the graph structure into the objective function.
In Sect.~\ref{subsec:formalization}, we stated that the strength of the link between $\mathcal{C}_p$ to $\mathcal{C}_q$ is proportional to the similarity of their respective \rev{prototypes}, $\mu_p$ and $\mu_q$.
We use the temporal-aware dissimilarity measure to assess the similarity of the \rev{prototypes}.
A low value for $|| \mu_p - \mu_q||_{TA}$ (high similarity) results in a high value for $a_{p,q}$ (powerful link between $\mathcal{C}_p$ and $\mathcal{C}_q$).
This translates into the second term of the objective function:
\begin{equation} \label{eq:term_2}
	T_{2} =  \sum_{\mu_p \in \mathcal{M}} \sum_{\substack{\mu_q \in \mathcal{M} \\ p \neq q}} a_{p,q}^2 || \mu_p - \mu_q||_{TA} \enspace.
	\vspace*{-0.01\textheight}
\end{equation}
$T_2$, together with  $\left(1-a_{p,q}^2 \right)$ in term $T_1$ (Eq.~\ref{eq:term_1}), assures the smooth passage for entities.
$T_1$ encourages successive observations to be assigned to clusters with a high value for $a_{p,q}$.
$T_2$ ensures that similar clusters have a strong link (high value for $a_{p,q}$).
Therefore, successive observations belonging to an entity are assigned to similar clusters, satisfying the \rev{third learning objective (\textbf{\textit{Obj. 3}})}.
The strength of the link between two clusters $\mathcal{C}_p$ and $\mathcal{C}_q$ is also dependent on the number of entities which present a transition from  $\mathcal{C}_p$ to $\mathcal{C}_q$.
We consider that an entity $\phi_l$ presents a transition between $\mathcal{C}_p$ to $\mathcal{C}_q$ ($\mathcal{C}_p \xrightarrow{\phi_l} \mathcal{C}_q$) if and only if two consecutive observations exist, associated with the given entity, where the first observation (ordered by their timestamp) is clustered under $\mathcal{C}_p$ and the second one under $\mathcal{C}_q$.
We define the intersection similarity measure between two phases, based on the normalized number of entities that present the transition between the two phases:
\begin{equation} \label{eq:intersection-measure}
	\vspace{-0.1in}
	inter_{\phi} (\mathcal{C}_p, \mathcal{C}_q) = 1 - \frac{|\{ \phi_l \in \Phi | \mathcal{C}_p \xrightarrow{\phi_l} \mathcal{C}_q \}|}{|\Phi|} \enspace,
\end{equation}
where $inter_{\phi} (\mathcal{C}_p,\mathcal{C}_q) \in [0,1]$ and a value of zero means that all entities present a transition from $\mathcal{C}_p$ to $\mathcal{C}_q$.
We construct the third term of the objective function:
\begin{equation} \label{eq:term_3}
	T_{3} =  \sum_{\mu_p \in \mathcal{M}} \sum_{\substack{\mu_q \in \mathcal{M} \\ p \neq q}} a_{p,q}^2 inter_{\phi}^2 (\mathcal{C}_p, \mathcal{C}_q) \enspace.
\end{equation}

We construct the objective function $\mathcal{J}$ as the weighted sum of the three terms in Eq.~\ref{eq:term_1}, \ref{eq:term_2} and~\ref{eq:term_3}:

\begin{align}
	\vspace*{-0.01\textheight}
	\mathcal{J} &= \lambda_1 T_1 + \lambda_2 T_2 + \lambda_3 T_3 = \nonumber \\
	&= \lambda_1  \sum_{\mu_p \in \mathcal{M}}\sum_{x_i \in \mathcal{C}_p} \left( || x_i - \mu_p||_{TA} + \sum_{\substack{x_k \in \mathcal{C}_q \\ q \neq p, \; x_i^\phi = x_k^\phi}}^{x_i^t < x_k^t} \beta * e^{-\frac{1}{2} \left( \frac{||x_i^t - x_k^t||}{\delta} \right)^2} \left(1-a_{p,q}^2 \right) \right)  \nonumber \\
	&+ \lambda_2 \sum_{\mu_p \in \mathcal{M}} \sum_{\substack{\mu_q \in \mathcal{M} \\ p \neq q}} a_{p,q}^2 || \mu_p - \mu_q||_{TA} + \lambda_3 \sum_{\mu_p \in \mathcal{M}} \sum_{\substack{\mu_q \in \mathcal{M} \\ p \neq q}} a_{p,q}^2 inter_{\phi}^2 (\mathcal{C}_p, \mathcal{C}_q) \enspace, \label{eq:objective-function}
\end{align}
where $\lambda_1, \lambda_2, \lambda_3 \in \mathbb{R^+}$ are parameters of the algorithm and they represent the weights of the different components.
They allow to fine-tune the impact of each of the stated objectives.
For example, if it is desirable to obtain evolution paths with small increments, in which the successive phases are very similar, it suffices to set $\lambda_2$ to a high value.
It would also result in paths with a larger number of phases than for a combination of weights featuring a lower $\lambda_2$.

The objective function $\mathcal{J}$ in Eq.~\ref{eq:objective-function} is artificially minimized if $a_{pq} = 0, \forall p, q$.
Given that $\mathcal{J} \in \mathbb{R^+}$, we constrain the 1-norm of the adjacency matrix $A$:
\begin{equation} \label{eq:constraint-adjacecy}
\vspace*{-0.01\textheight}
||A||_1 = 1 \Leftrightarrow	 \sum_{p=1}^{k} \sum_{q=1}^{k} a_{p,q} = 1 \enspace.
\end{equation}
This strategy allows to set high values of $a_{p,q}$ for pairs of clusters for which \linebreak i) $|| \mu_p - \mu_q||_{TA}$ is low (similar \rev{prototypes}) and ii) $inter_{\phi}^2 (\mathcal{C}_p, \mathcal{C}_q)$ is low (a high number of entities present a transition from $\mathcal{C}_p$ to $\mathcal{C}_q$).

\vspace*{-0.03\textheight}
\subsection{Optimizing the Objective Function. The \cluspath{} Algorithm.}
\label{subsec:cluspath-algo}
\vspace*{-0.01\textheight}

We minimize the objective function $\mathcal{J}$ using a \KMeans{}-like algorithm, by estimating the inner variables: i) the assignment of \rev{observations} to clusters, ii) the \rev{prototypes} of clusters and iii) the adjacency matrix.
\cluspath{} uses an iterative relocation strategy, at each step considering as fixed two of the variables (\textit{e.g.}, the observations assignment and the \rev{prototypes}) and analytically computing the values of the third (the adjacency matrix).

\textbf{Assignment of observations to clusters.}
For each observation $x_i$, \cluspath{} chooses a cluster $\mathcal{C}_p$ so that $\mathcal{J}$ is minimized.
Considering that $T_2$ and $T_3$ in Eq.~\ref{eq:objective-function} are not dependent on the assignment of observations to clusters, choosing a cluster for $x_i$ boils down to minimizing $T_1$:
\begin{equation} \label{eq:assign-update}
	\vspace*{-0.01\textheight}
	\textbf{best\_cluster}(x_i) = \underset{p = 1,2,...,k}{argmin} \left( || x_i - \mu_p||_{TA} + \sum_{\substack{x_k \in \mathcal{C}_q \\ q \neq p \\ x_i^\phi = x_k^\phi}}^{x_i^t < x_k^t} \beta * e^{-\frac{1}{2} \left( \frac{||x_i^t - x_k^t||}{\delta} \right)^2} \left(1-a_{p,q}^2 \right) \right)
\end{equation}
This guaranties that the contribution of $x_i$ to the value of $\mathcal{J}$ diminishes or stays constant.
Overall, this assures that $\mathcal{J}$ does not increase in the assignment phase.

\textbf{Recomputing the \rev{prototypes} of the clusters.}
\cluspath{} updates the \rev{prototypes} by recomputing one \rev{prototype} at a time (\textit{e.g.}, $\mu_j$), while considering all the other \rev{prototypes} fixed at their values from the previous iteration.
Each of the $k$ \rev{prototypes} $\mu_j \in \mathcal{M}$ is composed from a descriptive component and a temporal component: $\mu_j = (\mu_j^d, \mu_j^t)$.
Given that the subproblem of recomputing one \rev{prototype} is quadratic and no additional constraints are imposed on $\mu_j$, each of the two components is updated individually, by calculating the fixed point.
Given that only the first two terms of the function $\mathcal{J}$ are dependent of $\mu_j$, from Eq.~\ref{eq:objective-function} 
we obtain the \rev{prototype} update formulas (details of the complete calculations are in the Supplementary Materials (SM)~\cite{supplemental}):
\begin{align}
	\mu_j^d &= \frac{ \lambda_1  \sum_{x_i \in \mathcal{C}_j} x_i^d \left(1 - \gamma_t \frac{||x_i^t - \mu_j^t||^2}{\Delta t_{max}^{2}}\right) + \lambda_2 \sum_{\substack{\mu_p \in \mathcal{M} \\ p \neq j}} \mu_p^d \left(1 - \gamma_t \frac{||\mu_p^t - \mu_j^t||^2}{\Delta t_{max}^{2}} \right) (a_{j,p}^2 + a_{p,j}^2 )}{\lambda_1 \sum_{x_i \in \mathcal{C}_j} \left(1 - \gamma_t \frac{||x_i^t - \mu_j^t||^2}{\Delta t_{max}^{2}} \right) + \lambda_2 \sum_{\substack{\mu_p \in \mathcal{M} \\ p \neq j}} \left(1 - \gamma_t \frac{||\mu_p^t - \mu_j^t||^2}{\Delta t_{max}^{2}} \right) (a_{j,p}^2 + a_{p,j}^2)} \nonumber \\
	\mu_j^t &= \frac{ \lambda_1  \sum_{x_i \in \mathcal{C}_j} x_i^t \left(1 - \gamma_d \frac{||x_i^d - \mu_j^d||^2}{\Delta d_{max}^{2}}\right) + \lambda_2 \sum_{\substack{\mu_p \in \mathcal{M} \\ p \neq j}} \mu_p^t \left(1 - \gamma_d \frac{||\mu_p^d - \mu_j^d||^2}{\Delta d_{max}^{2}} \right) (a_{j,p}^2 + a_{p,j}^2 )}{\lambda_1 \sum_{x_i \in \mathcal{C}_j} \left(1 - \gamma_d \frac{||x_i^d - \mu_j^d||^2}{\Delta d_{max}^{2}} \right) + \lambda_2 \sum_{\substack{\mu_p \in \mathcal{M} \\ p \neq j}} \left(1 - \gamma_d \frac{||\mu_p^d - \mu_j^d||^2}{\Delta d_{max}^{2}} \right) (a_{j,p}^2 + a_{p,j}^2)} \enspace. \label{eq:centroid-update}
\end{align}

Similarly to \KMeans{}, the \rev{prototypes} computed in Eq.~\ref{eq:centroid-update} represent prototypes of their clusters.
Unlike \KMeans{}, $\mu_j$ the \rev{prototype} of a cluster $\mathcal{C}_j$ is not only the average of the observations regrouped under $\mathcal{C}_j$, but it is also influenced by the \rev{prototypes} of the other clusters linked to $\mathcal{C}_j$.
Moreover, it represents the average of the assigned observations on both the temporal dimension, as well as on the descriptive dimension, with the two dimensions weighting each other.
\textit{E.g.}, temporally central observations weight more in the calculation of $\mu_j^d$, the descriptive component of the new \rev{prototype}.
Furthermore, the contributions of the other \rev{prototypes} ($\mu_p$) are weighted by the strength of the link between $\mathcal{C}_p$ and $\mathcal{C}_j$, the cluster of the currently recomputed \rev{prototype} $\mu_j$.
\rev{Inferring the relations between clusters simultaneously with the clustering represents one of the advantages of \cluspath{}}: the relations between clusters can also influence their content.
The orientation of the link is not important, since both $a_{p,j}$ and $a_{j,p}$ appear in the update formula.

\textbf{Updating the adjacency matrix.}
Similarly to recomputing the \rev{prototypes}, the update of the adjacency matrix is a quadratic problem.
The difference is that the additional constraint in Eq.~\ref{eq:constraint-adjacecy} is imposed, in order to avoid a trivial null solution.
A typical solution to optimizing an equality constraint problem is to use a Lagrange multiplier on the constraint, similarly to~\cite{DUN73}.
Consequently, the update formula for the adjacency matrix is the solution of the following:
\begin{equation} \label{eq:adj-derivates}
	\vspace*{-0.01\textheight}
	\frac{\partial \mathcal{J}^*}{\partial a_{r,s}} = 0, \text{ where } \mathcal{J}^* = \mathcal{J} - \lambda \left( \sum_{p=1}^k \sum_{q=1}^k a_{p,q} - 1 \right) \enspace.
\end{equation}
By \rev{computing the point in which} the derivative of $\mathcal{J}^*$ \rev{is null}, we obtain the adjacency matrix update formulas (for complete calculations see SM~\cite{supplemental}):
\begin{align} 
	a_{r,s} =& \frac{1}{K_{r,s} \sum_{p=1}^{k} \sum_{q=1}^{k} \frac{1}{K_{p,q}} } \enspace, \nonumber \\
	&\text{with } K_{r,s} =  -\lambda_1 \; pen(\mathcal{C}_r \xrightarrow{\phi} \mathcal{C}_s) + \lambda_2 || \mu_r - \mu_s ||_{TA} + \lambda_3 \; inter_{\phi}^2 (\mathcal{C}_p, \mathcal{C}_q) \enspace, \nonumber \\
	&\text{and } pen(\mathcal{C}_r \xrightarrow{\phi} \mathcal{C}_s) = \sum_{x_i \in \mathcal{C}_r} \sum_{\substack{x_k \in \mathcal{C}_s \\ x_i^\phi = x_k^\phi}}^{x_i^t < x_k^t} \left( \beta * e^{-\frac{1}{2} \left( \frac{||x_i^t - x_k^t||}{\delta} \right)^2} \right) \enspace. \label{eq:adjacency-update}
	\vspace*{-0.01\textheight}
\end{align}

The $pen(\mathcal{C}_r \xrightarrow{\phi} \mathcal{C}_s)$ function in Eq.~\ref{eq:adjacency-update} is the influence that the contiguous penalty defined in Eq.~\ref{eq:penalty-function} has on the adjacency update process.
Intuitively, this mechanism allows to take into account the transitions of entities between phases, when computing the adjacency matrix.
If many entities present transitions between clusters $\mathcal{C}_r$ and $\mathcal{C}_s$, then $a_{r,s}$ will be recomputed at a large value.
This, in turn, allows in the subsequent iteration to lower the contiguity penalty, by lowering the $1 - a_{r,s}^2$ term in Eq.~\ref{eq:penalty-function}.
The similarity between the \rev{prototypes} $\mu_r$ and $\mu_s$ and the number of entities presenting a transition between $\mathcal{C}_r$ and $\mathcal{C}_s$ also impact the adjacency matrix update through $K_{r,s}$.

\textbf{The \cluspath{} Algorithm.}
The outline of \cluspath{} is given in Algorithm~\ref{algo:proposed-algo}.
\cluspath{} seeks to minimize $\mathcal{J}$ by iterating an assignment phase, a \rev{prototype} update phase and an adjacency matrix update phase until the partition does not change between two iterations.
Aside from $k$ (the number of clusters), \cluspath{} uses six parameters: $\alpha$, $\beta$, $\delta$, $\lambda_1, \lambda_2, \lambda_3$.
In Sect.~\ref{subsec:choose-params-evolutionary}, we discuss a technique for tuning these parameters to an unseen dataset.
A random subset of observations $x_l \in \mathcal{X}, 1 = 1,2,..k$ can be used as $\mathcal{M}^{(0)}$, the initial set of \rev{prototypes}.
The \textbf{up\_cluster} and \textbf{up\_adjacency} are two functions used to recompute, respectively, the \rev{prototypes} and the adjacency matrix, as shown before in this section.
Similarly to other gradient descent algorithms, \cluspath{} may present the usual shortcomings, such as converging to a local optima or slow convergence speeds near the minimum. 
While the convergence speed of \cluspath{} has not been theoretically studied, the experiments in Section~\ref{sec:experiments} show that the optimization process practically stops in a few steps.

\begin{algorithm}[tb]                     
\caption{Outline of the \cluspath{} Algorithm.}
\label{algo:proposed-algo} 
\small
\DontPrintSemicolon
	
\KwData{observations $x_i \in \mathcal{X}$, set of initial \rev{prototypes} $\mathcal{M}^{(0)}$}
\KwResult{\rev{prototypes} $\mu_j, j = 1,..,k$, clusters $\mathcal{C}_j, j = 1,..,k$, adjacency matrix $A$}
\textbf{Parameters: } number of clusters $k$, $\alpha$, $\beta$, $\delta$, $\lambda_1, \lambda_2, \lambda_3$ 

\SetSideCommentLeft{\emph{  // \rev{adjacency matrix initialization} \\}}
$a_{i,j}^{(0)} = 0, \;\; \forall i,j = 1,2,..,k$\;
$ iter \gets 0$  \;
$\mathcal{P}^{(iter)} \gets \emptyset$ 
\SetSideCommentRight{\emph{\hspace{5mm} //set of \rev{phases}}} \;

\Repeat {$\mathcal{C}_j^{(iter)} = \mathcal{C}_j^{(iter-1)}, \forall j = 1,2,..,k$} {
	$iter \gets iter + 1$ \;
	\For {$j = 1,2,...,k $} {
		$\mathcal{C}_j^{(iter)} \gets \emptyset$ \;
	}
	
	\SetSideCommentLeft{\emph{  // \rev{S1. observation assignment to phases}}}\;
	\For { $x_i \in \mathcal{X}$ } {
		$ j = $ \textbf{best\_cluster}($x_i$, $\mathcal{X}$, $\mathcal{M}^{(iter-1)}$, $\mathcal{P}^{(iter-1)}$, $A^{(iter-1)}$) \;
		$\mathcal{C}_j^{(iter)} = \mathcal{C}_j^{(iter)} \cup \{ x_i \}$  \;
	}
	
	\SetSideCommentLeft{\emph{  // \rev{S2. update prototypes}}}\;
	\For {$j = 1,2,...,k $} {
		$ (\mu_j^{d, (iter)}, \mu_j^{t, (iter)}) \gets $ \textbf{up\_\rev{prototype}}($j$, $\mathcal{X}$, $\mathcal{M}^{(iter-1)}$, $\mathcal{P}^{(iter-1)}$,  $A^{(iter-1)}$)\;
	}

	\SetSideCommentLeft{\emph{  // \rev{S3. update adjacency matrix}}}\;
	$A^{(iter)} \gets $ \textbf{up\_adjacency}($\mathcal{X}$, $\mathcal{M}^{(iter-1)}$, $\mathcal{P}^{(iter-1)}$)\;

	$\mathcal{M}^{(iter)} \gets \{\mu_j^{(iter)} | \; j = 1,2,..,k\}$\;
	$\mathcal{P}^{(iter)} \gets \{\mathcal{C}_j^{(iter)} | \; j = 1,2,..,k\}$\; 
}
\end{algorithm}

\textbf{Algorithm complexity.}
\rev{
We denote by $T(x)$ the time complexity of subroutine $x$.
From Algorithm~\ref{algo:proposed-algo} is results $T(\cluspath{}) = T(S1) + T(S2) + T(S3)$.
If $p$ is the number of entities and $N$ is the number of observations associated with each entity, then $n = p \times N$ is the total number of observations.
We assume that $k << n$.
We compute $T(S1)$ as $n T(\textbf{best\_cluster}) = pN T(\textbf{best\_cluster})$.
Due to the penalty term in Eq.~\ref{eq:assign-update}, $T(\textbf{best\_cluster}) = \mathcal{O}(kN)$, therefore $T(S1) = \mathcal{O}(pN^2k)$.
The complexity of updating centroids (\emph{S2}) is $k T(\textbf{up\_prototype})$.
From Eq.~\ref{eq:centroid-update} results $T(\textbf{up\_prototype}) = 2 (\mathcal{O}(n) + \mathcal{O}(k)) = \mathcal{O}(n)$, therefore $T(S2) = \mathcal{O}(pNk))$.
Lastly, $T(S3) = T(\textbf{up\_adjacency}) = k^2 T(K_{r,s})$, which can be obtained from Eq.~\ref{eq:adjacency-update}.
Computing $a_{r,s}$ is dependent on computing $K_{r,s}$, which need be computed only once per iteration.
$T(K_{r,s}) = T \left( inter_{\phi}^2 (\mathcal{C}_p, \mathcal{C}_q) \right) + T \left( pen(\mathcal{C}_r \xrightarrow{\phi} \mathcal{C}_s) \right)$.
From Eq.~\ref{eq:intersection-measure} results $T \left( inter_{\phi}^2 (\mathcal{C}_p, \mathcal{C}_q) \right) = \mathcal{O}(pN)$ (since we need to iterate only once through the observations of an entity to detect transitions).
Furthermore, $T \left( pen(\mathcal{C}_r \xrightarrow{\phi} \mathcal{C}_s) \right) = \mathcal{O}(nN) = \mathcal{O}(pN^2)$.
Consequently, $T(K_{r,s}) = \mathcal{O}(pN^2) \Rightarrow T(S3) = \mathcal{O}(pN^2k^2)$.
This amounts to a complexity of \cluspath{} of $\mathcal{O}(pN^{2}k^2)$, which is well adapted to Social Science and Humanities datasets, where often a large number of entities is studied over a relatively short period of time ($p > N$).
}

\textbf{Heuristics for Displaying the Constructed Graph.}
The adjacency matrix $A = (a_{i,j})$ shows the strength of the link between each pair of clusters.
$a_{i,j} \in \mathbb{R}, i,j = 1,2,..,k$.
To display the relation between clusters as a graph, we construct a binary matrix $A^*$, using a simple heuristic: a threshold $\lambda$ is chosen so that only the $k-1$ scores are retained. 
This value is chosen to favor a tree structure, even if a tree cannot be guaranteed given the structure of the graph (\textit{i.e.}, some nodes may be central, with many connections, while others are marginal or even unconnected).
All arcs having the selected scores are plotted, consequently, more than $k-1$ arcs might be used.
$a^*_{i,j} = 1$ \textit{iff} $a_{i,j} > \lambda$, and $a^*_{i,j} = 0$ \textit{iff} $a_{i,j} \leq \lambda$.
\textcolor{black}{
Unconnected nodes are eliminated, as they are considered isolated evolution phases and, therefore, not suitable for an evolution path.
Given the adjacency matrix update formula in Eq.~\ref{eq:adjacency-update}, $a_{i,j}$ is low when entities do not present transitions between $\mathcal{C}_i$ and $\mathcal{C}_j$ and $\mu_i$ and $\mu_j$ are very dissimilar.
Therefore, entities which find themselves in these unconnected phases do not transition into other phases and are dissimilar to the others.
In other words, they act as outliers of the typical evolution path.
}

\vspace*{-0.03\textheight}
\subsection{Automatically Tuning the Parameters Using Evolutionary Algorithms}
\label{subsec:choose-params-evolutionary}
\vspace*{-0.01\textheight}

\cluspath{} uses six parameters: $\alpha$, $\beta$, $\delta$, $\lambda_1, \lambda_2, \lambda_3$, which can prove challenging to tune, especially on new, unseen datasets.
Datasets issued from different domains, like the ones presented in Sect.~\ref{sec:experiments}, may have different requirements for the ratio between the descriptive and temporal dimensions (the $\alpha$ parameter)  or the smoothness of the evolution path (the $\lambda_2$ parameter).
We provide an optional method for automatically tunning \rev{the parameters of \cluspath{}}, at the expense of repeated runs of the algorithm.
Using an evolutionary technique, we optimize over the 6-dimensional space of the parameters in order to find a solution which provides a balance between the four opposite measures used to evaluate its output.

\textbf{Evaluating a partition.}
We use the classical Information Theory measures to numerically assess the four objectives defined in Sect.~\ref{subsec:formalization}.
The first three objectives are evaluated using measures proposed by~\cite{RIZ12}.
The coherence of the obtained partition in the descriptive and temporal dimensions are \rev{measured} using the classical variance on, respectively, the multidimensional component (\textit{MDvar} measure) and the temporal component (\textit{Tvar} measure).
\rev{We compute the variance as the mean within-cluster dissimilarity between observations and their associated prototype.}
The contiguous segmentation of the series of observations corresponding to an entity is measured using a penalized Shannon entropy (\textit{ShaP}), in which a sequentiality component is added by weighting the value of the classical entropy by a penalty factor depending on the number of continuous segments in the series of each entity:
\begin{equation*}
	\vspace*{-0.01\textheight}
	\rev{ShaP = \frac{1}{|\Phi|}  \sum_{\phi \in \Phi} \sum_{i=1}^k \left[ -p_{\phi}(\mathcal{C}_i)\log_2(p_{\phi}(\mathcal{C}_i)) \right] \left( 1 + \frac{n_{ch} - n_{min}}{N - 1}\right), \enspace p_{\phi}(\mathcal{C}_i) = \sum_{\substack{x_j \in \mathcal{C}_i \\ x_j^\phi = \phi}} \frac{1}{N}}
%
\end{equation*}
where $n_{ch}$ is the number of changes in the cluster assignment series of an entity, $n_{min}$ is the minimal required number of changes and $N$ is the number of observations for an entity.
We add a forth measure (\textit{SPass}) to assess the smooth passage of entities along an evolution path  measure:
\begin{equation*}
	\vspace*{-0.01\textheight}
	\textit{SPass} = \sum_{\phi \in \Phi} \sum ^{\rev{i,j \in 1,\ldots, k}} _{\mathcal{C}_i \xrightarrow{\phi}  \mathcal{C}_j} \frac{|| \mu_i - \mu_j||_{TA}}{n_{ch}} \enspace.
\end{equation*}
which measures the average temporal-aware dissimilarity between successive phases.
Therefore, to each solution constructed by \cluspath{} corresponds a point in the four-dimensional space of the measures: (\textit{MDvar}, \textit{Tvar}, \textit{ShaP}, \textit{SPass})

\textbf{Defining a balanced solution. The parameter tuning heuristic.}
The four objectives defined in Sect.~\ref{subsec:formalization} and the four corresponding measures defined here above are contradicting; completely fulfilling them at the same time is not possible.
To acquire a trade-off between the mutually contradicting objectives, \rev{we pose the problem of automatically tuning the parameters of \cluspath{} as an optimization problem. 
We consider the parameters of \cluspath{} as internal variables over which the
search is performed and we evaluate using the four evaluation criteria.
To chose a balanced solution,} we use the concept of Pareto optimality~\cite{SAW85}, originally developed in economics. 
Given a set of solutions, a given solution is considered to be Pareto optimal if there exists no other that, simultaneously, \rev{is better on all objectives}. 
The set of Pareto optimal solutions form the Pareto front.
Consequently, no single optimum can be constructed, but rather a class of optima, depending on the ratio between the objectives.
With no \textit{a priori} information, selecting the point on the Pareto front closest (in terms of Euclidean distance) to the ideal point provides a good compromise between the different objectives.
All measures need to be minimized, therefore the ideal point is $(0, 0, 0, 0)$.
By default, we use no weights on the dimensions when calculating the distance from a solution to this point.
Before computing the Euclidean distance to the ideal point, the values of all measures are normalized, in order to receive equal importance.
\rev{Having the Pareto constructed allows using another methodologies for selecting the ``good trade-off'' without re-executing the lengthy optimization procedure.}

\textbf{Approximating the Pareto frontier using evolutionary algorithms.}
The heuristic we propose is, for a given dataset, to construct the Pareto front in the measures space, which is the 4-dimensional envelope of all the possible compromises.
The solution closest to the ideal point is chosen as the ``best'' compromise and its corresponding parameter values are presented as the tuned values for the particular dataset.
We formulate the problem of parameter tunning as a multi-objective optimization problem, optimizing in the 4-dimensional space of evaluation measures, with the parameters of \cluspath{} serving as internal variables over which the search is performed.
Solving multiobjective optimization problems using evolutionary algorithms (MOEAs) has been investigated by many authors~\cite{DEB02,HAL06,KAF11,MIH14,ZIT01}. 
Pareto dominance based MOEAs such as NSGAII~\cite{DEB02}, SPEA2~\cite{ZIT01} and HEMH~\cite{KAF11} have been dominantly used in the recent studies. 
In multiobjective optimization, the set of Pareto optimal solutions is approximated using a large number of non-dominated points. 
MOEAs operates on individuals of an initial population to generate the individuals of the population of the next generation. 
The new population is generated by applying some processes of selection, recombination and mutation.

\textbf{Our technique}
The genome of each individual is a vector composed of the six parameters of \cluspath{}: $\alpha$, $\beta$, $\delta$, $\lambda_1, \lambda_2, \lambda_3$.
For each individual, \cluspath{} is executed with the parameters in its genome, a solution is obtained and it is evaluated.
Therefore, to each individual corresponds a point in the 4-dimensional space of the measures.
The size \rev{of each evolutionary population} is fixed to 100.
\rev{We use the Pareto dominance to evaluate the fitness function: the number of individuals which dominate the given individual.
An elitist selection is used to filter the population: }all the non-dominated solutions and 10\% of the dominated solutions are promoted in the next generation, while the rest are removed.
We use two operators to construct, based on the selected elite, new solutions until the population reaches the nominal size.
We duplicate and mutate 5\% of the survivors.
The mutations affect one or two parameters in the genome, which are set to a random value in their domain of definition.
New \rev{offsprings} are generated through a \textit{Path-Relinking} strategy: two parents are selected from the survivors and the values for the newly generated \rev{individuals} are averaged means of the values of the parents.
The weights are randomly generated.
The process is iterated until all the constructed solutions are Pareto non-dominated or until a maximum number of generations (100) is reached.
We choose as the ``best'' solution, the individual in the last generation \rev{which is the} closest to the ideal point.
Given the elitist strategy, non-dominated solutions always survive into successive generations.

\textbf{The complexity of the evolutionary heuristic.}
\rev{For a given partition constructed by \cluspath{}, we have $T(MDvar) = T(Tvar) = T(ShaP) = \mathcal{O}(pNk)$ and $T(SPass) = \mathcal{O}(pN)$.
Given $m$ the number of individuals in each evolutionary generation, evaluating all the individuals in a generation takes $\mathcal{O}(mpNk)$.
Calculating the fitness function at each generation has the complexity $\mathcal{O}(m^{2})$.}
Sorting the individuals by their fitness is done in $\mathcal{O}(m \times log(m))$.
\cluspath{} is executed for each individual, therefore a complexity of $m \times \mathcal{O}(pN^{2}k^2) = \mathcal{O}(mpN^{2}k^2)$ (cf. complexity of \cluspath{} calculated in Sect.~\ref{subsec:cluspath-algo}).
Therefore, the complexity of each evolutionary generation is $\rev{\mathcal{O}(mpNk) +} \mathcal{O}(m^2) + \mathcal{O}(mpN^{2}k^2) = \mathcal{O}(mpN^{2}k^2)$, considering that $m << pN^2k^2$.
This results in a total complexity of the evolutionary heuristic of $\mathcal{O}(ng_{max} \times mpN^{2}k^2)$, where $ng_{max}$ is the maximum number of generations.

Note that the calculated complexity is the worst case scenario.
In practice, the elitist technique speeds up the computation considerably, since the solutions promoted from the previous generation do not require a re-execution of the \cluspath{} algorithm.
Furthermore, the elitist technique speeds up convergence and reduces the actual number of required generations.
Finally, evolutionary algorithms are very well adapted for massive parallelization.
Each individual execution of \cluspath{} is independent of the others and all the individuals in a generation can be computed in parallel, provided that the heuristic is ran on a machine with at least $m$ execution cores.
In our experiments in Section~\ref{sec:experiments}, the evolutionary algorithm proved to be only $6.12$ times slower than \cluspath{} (standard deviation of $0,46$ over 200 runs, all runtimes and hardware specs in SM~\cite{supplemental}).

\vspace*{-0.03\textheight}
\section{Experiments}
\label{sec:experiments}
\vspace*{-0.01\textheight}


The experiments are conducted on two real-life datasets, one issued from political sciences: \textit{Comparative Political Data Set I} (\texttt{CPDS1})~\cite{ARM11}  and the second containing financial and accounting data: \textit{European Companies} (\texttt{EC})~\cite{Siddiqui2012}.
\texttt{CPDS1} is a collection of political and institutional data, which consists of annual data for 23 democratic countries for the period from 1960 to 2009.
The dataset was cleaned by removing redundant variables (\textit{e.g.} country identifier and postal code), resulting in each country being described using 207 political, demographic, social and economic variables.
\rev{The corpus was preprocessed by removing entity-specific, time-invariant bias from the data.
For every attribute, we compute its mean for each entity.
For every pair (attribute, observation) we substract from the attribute value the mean of the corresponding entity.}
The obtained dataset\footnote{Download pretreated version of the \texttt{CPDS1} dataset here: \url{http://goo.gl/17ihsf}} is under the form of triples $(country, year, description)$.
The \texttt{EC} dataset describes the activity of 836 companies over a period of 5 years (2003-2007), using 7 economic variables.
The dataset is preprocessed similarly to \texttt{CPDS1}, by removing the entity specific means from each variable.
\rev{Note that the \texttt{EC} dataset is very different from \texttt{CPDS1}, in the sense that a consistently larger number of entities are studied over a short time span (5 timepoints for \texttt{EC} vs. 50 for \texttt{CPDS1}) and are described using few attributes (7 for \texttt{EC} vs. 207 for \texttt{CPDS1}).
As the experiments in the following sections show, this leads to less diverse evolution phases and shorter evolution paths.}
All attributes are normalized prior to the execution of the algorithms (to avoid setting artificial weights), whatsoever the results presented in this section are non-normalized, in order to appreciate the amplitude of the time-varying component.

\vspace*{-0.02\textheight}
\subsection{Choosing the Best Possible Solution Using an Evolutionary Algorithm}
\label{subsec:chose-solution-evolutionary}
\vspace*{-0.01\textheight}

\begin{figure}[tb]
	\centering
	\includegraphics[width=0.99\textwidth]{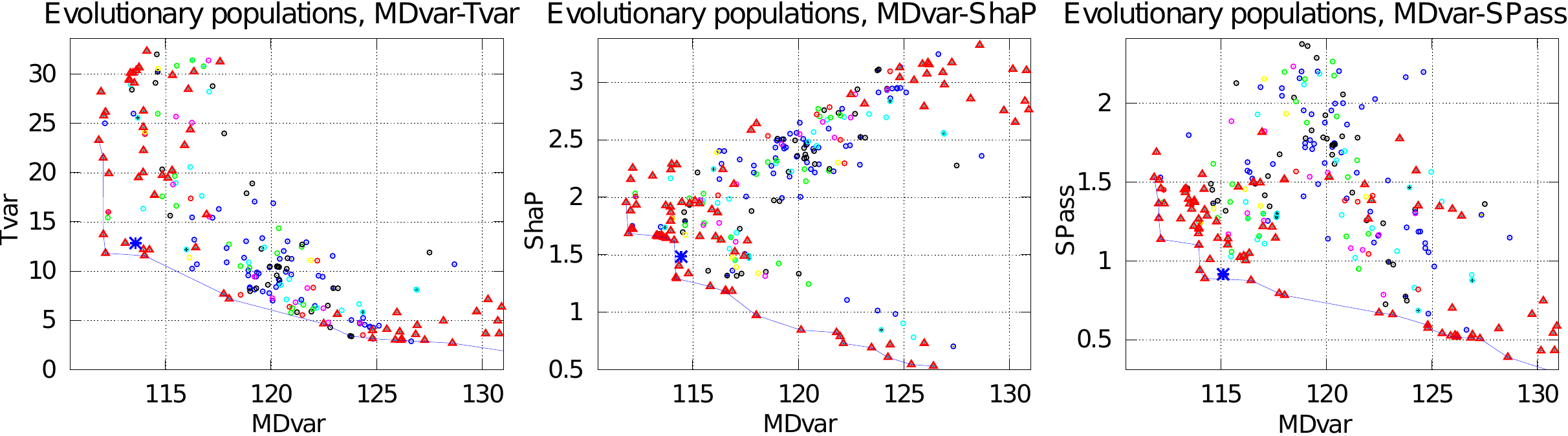}
	\hfill
	\includegraphics[width=0.99\textwidth]{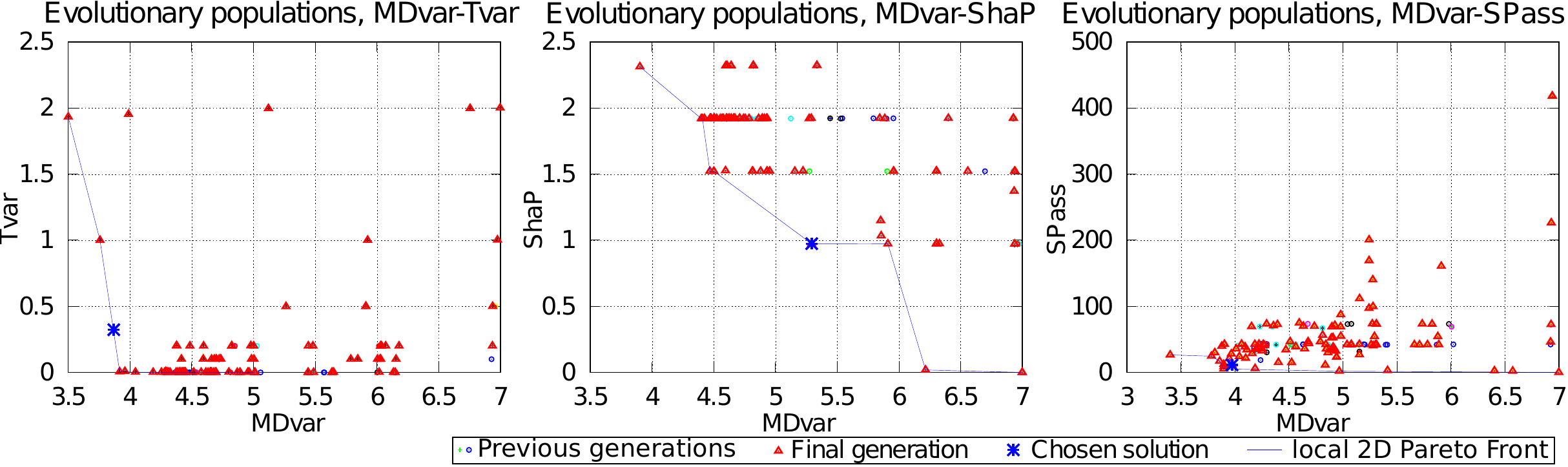}
	\caption{Typical example of execution of the evolutionary algorithm on \texttt{CPDS1} (top row) and on \texttt{EC} (bottom row). The obtained 4-dimensional Pareto front is projected onto the (\textit{MDvar}, \textit{Tvar}) space (left), (\textit{MDvar}, \textit{ShaP}) space (middle) and (\textit{MDvar}, \textit{SPass}) space (right).}
	\label{fig:pareto-criteria-all-projections}
\end{figure}

Throughout our experiments, \rev{the parameters of \cluspath{}} (except for $k$, the number of clusters) are chosen using the evolutionary heuristic described in Sect.~\ref{subsec:choose-params-evolutionary}.
Being a clustering algorithm, \cluspath{} suffers from the classical drawback of selecting the number of evolution phases.
Setting $k$ to large values \rev{results in clusters containing less observations}, more compact in description and time, adapted for detecting granular evolutions.
It also result in longer evolution paths.
Setting $k$ to a low number results in more general evolution phases, more adapted to detecting more general evolutions.
Fig.~\ref{fig:pareto-criteria-all-projections} presents a typical execution of the evolutionary algorithm, together with the obtained Pareto front and the chosen solution.
Each individual in each evolutionary generation is associated with a point in the 4-dimensional space of the measures: (\textit{MDvar}, \textit{Tvar}, \textit{ShaP}, \textit{SPass}).
For presentation reasons, we project the 4-dimensional space onto 2-dimensional spaces, by selecting pairs of measures.
In Fig.~\ref{fig:pareto-criteria-all-projections}, some of the points on the 4-dimensional Pareto front (indicated by the red triangles) seem Pareto dominated in the 2-dimensional spaces.
The 2-dimensional graphics present only projections and, as discussed in Sect.~\ref{subsec:choose-params-evolutionary}, optimizing multiple criteria means finding a compromise which rarely yields the best results on the individual criteria.
Similarly, the chosen global solution is not necessarily the optimum in each of the 2-dimensional spaces.
Whatsoever, Fig.~\ref{fig:pareto-criteria-all-projections} clearly shows that the chosen point is never too far from the local Pareto fronts and, thus, it provides a good trade-off.
On \texttt{EC}, the \textit{Tvar} measure (bottom row, left and center graphics) presents levels.
This is due to the very limited temporal extent of the dataset (5 years, \textit{i.e.}, 5 data points), which in turn limits the number of values that can be taken by the temporal variance.

\textbf{Runtime} 
An average run of \cluspath{} on \texttt{CPDS1} takes $122.81s$ ($stdev = 11.37$) on our \rev{24 cores Intel(R) Xeon(R) E5-2430} machine (see SM~\cite{supplemental} for \rev{full} machine specs), while the evolutionary technique takes $750.69s$ ($stdev = 53.8$).
This makes an execution of the evolutionary technique as long as roughly six sequential executions of \cluspath{}.

\begin{figure}[tb]
\centering
	\subfloat[] {
		\includegraphics[height=0.2\textheight]{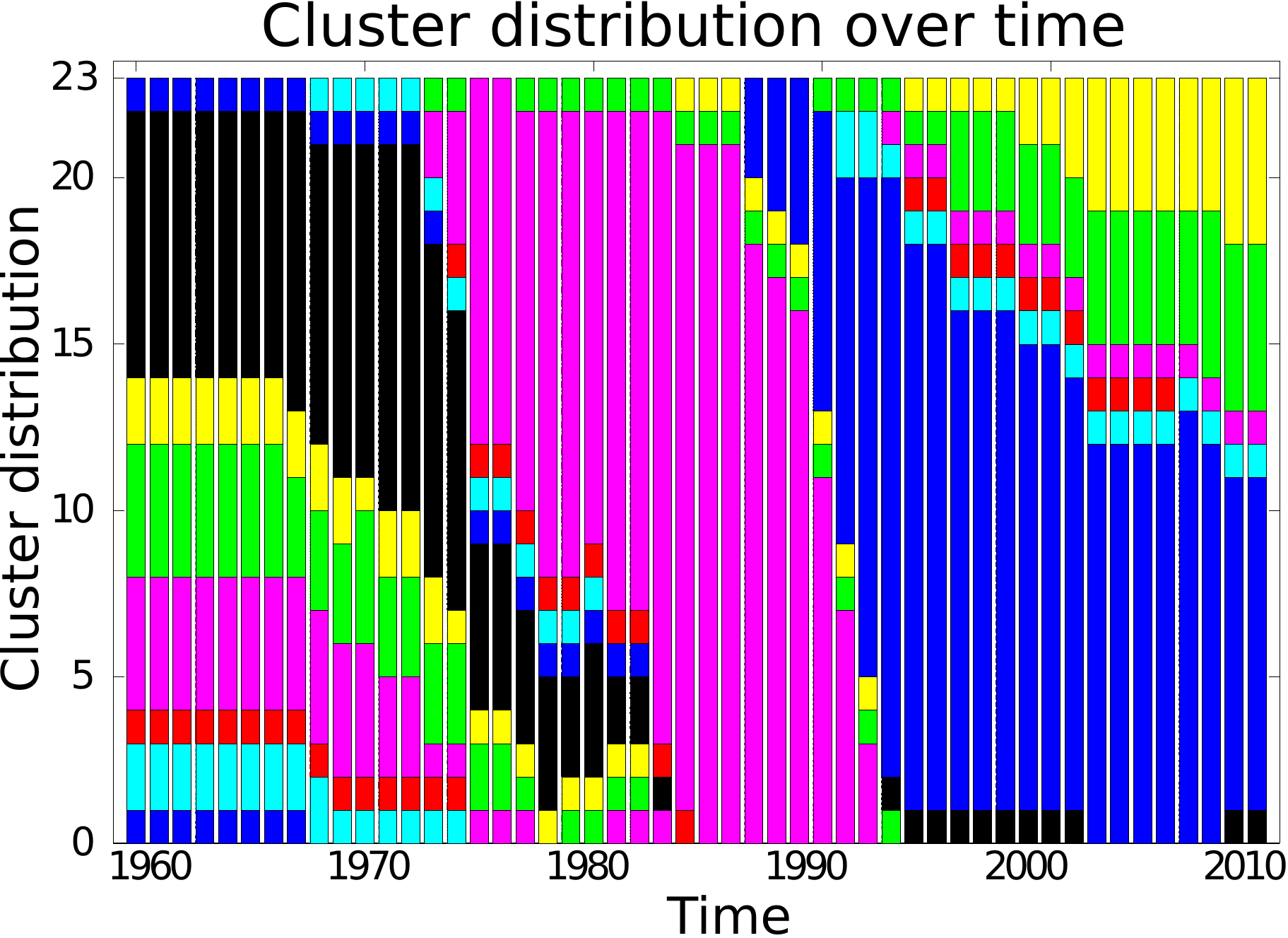}%
		\label{subfig:cluspath-clusters-vs-time}
	}
	\hspace{1cm}
	\subfloat[]{
		\includegraphics[height=0.2\textheight]{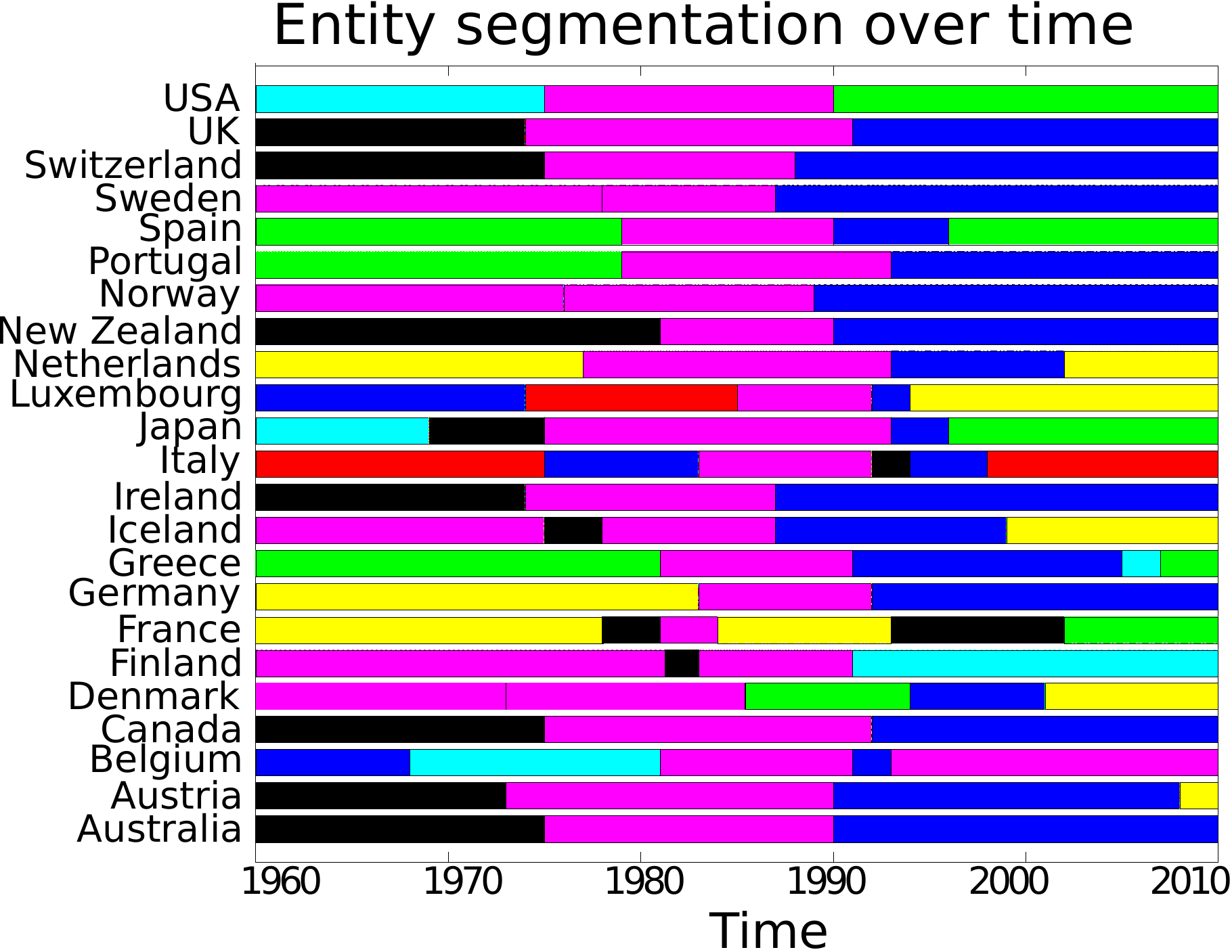}%
		\label{subfig:cluspath-entity-segmentation-bars}
	}
	\hfill
	\subfloat[]{
		\includegraphics[width=0.8\textwidth]{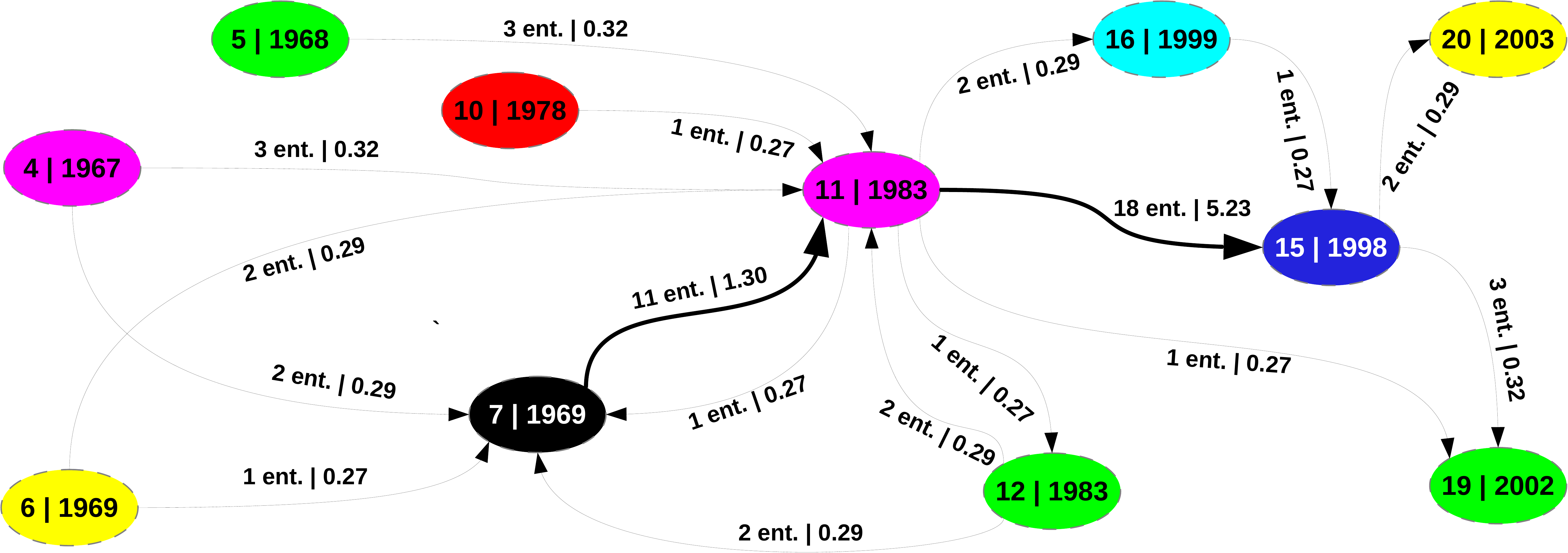}%
		\label{subfig:cluspath-graph}
	}

	\caption{Typical evolution phases constructed by \cluspath{} on  \texttt{CPDS1}, with 20 clusters. Number of entities in each phase per year (a), segmentation of entities over phases (b) and the phase evolution graph (c)}
	\label{fig:cpds1-qualitative-evaluation}
\end{figure}

\textbf{Qualitative Results}
Fig.~\ref{fig:cpds1-qualitative-evaluation} shows the typical evolution paths constructed by \cluspath{} on \texttt{CPDS1}, when asked for 20 clusters.
The heuristic used for displaying the constructed graph eliminates \rev{unconnected phases}.
Fig.~\ref{subfig:cluspath-clusters-vs-time} shows how many countries belong in a certain cluster for each year.
Clusters $\mathcal{C}_7$ (black), $\mathcal{C}_{11}$ (magenta) and $\mathcal{C}_{15}$ (blue) contain most of the observations, suggesting that the path $\mathcal{C}_7 \longrightarrow \mathcal{C}_{11} \longrightarrow \mathcal{C}_{15}$ is a typical evolution path followed by most entities.
The meaning of each constructed cluster unravels when studying the segmentation of countries over clusters, in Fig.~\ref{subfig:cluspath-entity-segmentation-bars}, as well as the proposed graph structure in Fig.~\ref{subfig:cluspath-graph}.
For example, the succession $\mathcal{C}_5 \longrightarrow \mathcal{C}_{11}$ is followed by Spain, Portugal and Greece at the beginning of their evolution.
Historically, this coincides with the non-democratic regimes in those countries (Franco's dictatorship in Spain, the ``Regime of the Colonels'' in Greece).
Likewise, the succession $\mathcal{C}_4 \longrightarrow \mathcal{C}_{11}$ (and the slightly longer $\mathcal{C}_4 \longrightarrow \mathcal{C}_7 \longrightarrow \mathcal{C}_{11}$) is present for countries like Denmark, Finland, Iceland, Norway and Sweden.
This evolution path maps onto the ``Swedish Social and Economical Model'' of the Nordic countries.	
\rev{For completeness reasons, we present in Fig.~\ref{fig:TDCKMeans-graph}, the graph of phases constructed \textit{a posteriori} by the extensions of \TDCKMeans{}, proposed in~\cite{Rizoiu2014b}.
Visibly, the constructed paths are longer (most transitions are followed by a single entity) and more difficult to interpret.
Furthermore, the entities evolution through the graph is not constrained during the clustering, which results in a significant number of \emph{backwards links} from phases with a higher timestamp to phases with a lower timestamp.
Overall, the graph generated by \cluspath{} is more synthetic and easier to interpret, since it contains less specific arcs (followed by a single entity) and less backward links.
This makes is more adapted for our application: identifying typical evolution paths.
}
%
%

\begin{figure}[tb]
\centering
	\subfloat[] {
		\includegraphics[height=0.18\textheight]{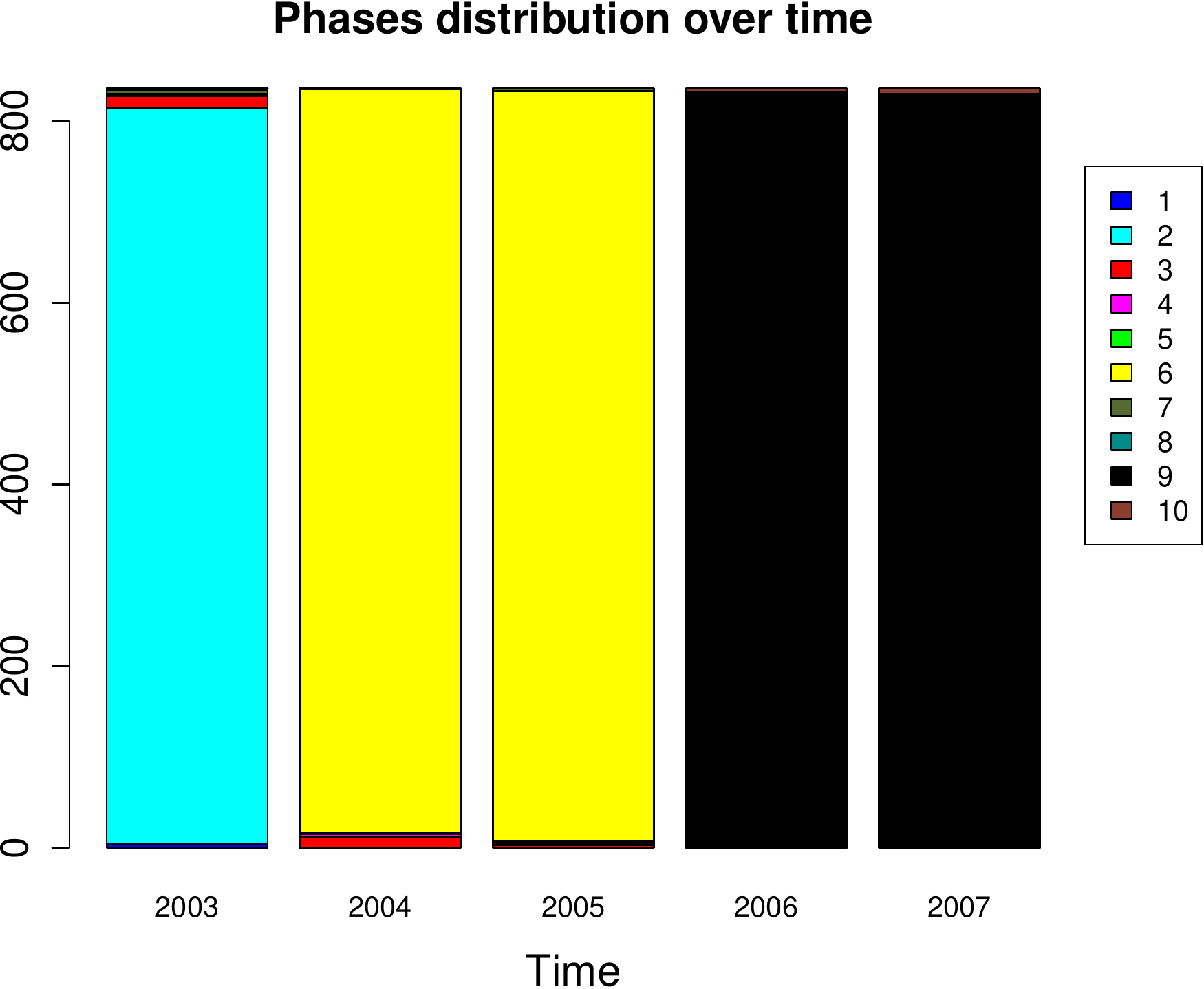}%
		\label{subfig:ec-cluspath-clusters-vs-time}
	}
	\hfill
	\subfloat[]{
		\includegraphics[height=0.18\textheight]{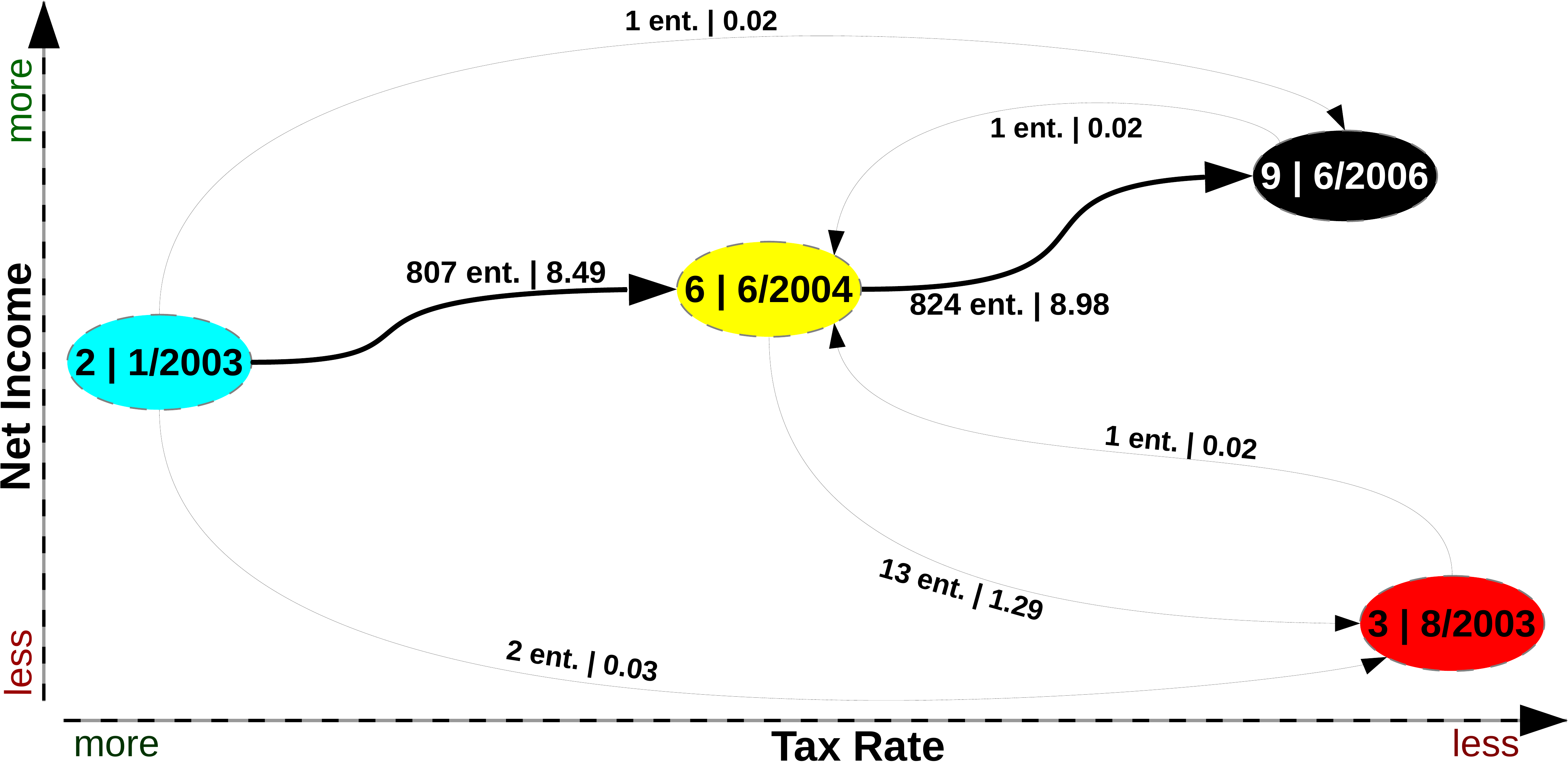}%
		\label{subfig:ec-cluspath-evoltution-graph}
	}

	\caption{Typical evolution phases constructed by \cluspath{} on \texttt{EC}, with 10 clusters. Number of entities in each phase per year (a) the evolution graph projected in the space \textit{NetIncome}/\textit{TaxRate} (b)}
	\label{fig:ec-qualitative-evaluation}
\end{figure}

On the \texttt{EC} dataset, \cluspath{} is executed with 10 clusters, because of the considerably shorter temporal extent of the dataset.
Figure~\ref{subfig:ec-cluspath-clusters-vs-time} shows that the phases $\mathcal{C}_2$, $\mathcal{C}_3$, $\mathcal{C}_6$ and $\mathcal{C}_9$ are the predominant evolution phases, with the evolution $ \mathcal{C}_2 \longrightarrow \mathcal{C}_6 \longrightarrow \mathcal{C}_9$ being the typical evolution in the population of companies.
The descriptions of these phases are given in Table~\ref{tab:centroids-EC}.
Considering that the dataset was preprocessed to remove entity-specific values, the negative and positive values in Table~\ref{tab:centroids-EC} indicate negative, respectively positive, tendencies.
For example, phase $\mathcal{C}_3$ is a crisis phase, in which companies reduce slightly their debt, while considerably reducing their revenues and income.

Figure~\ref{subfig:ec-cluspath-evoltution-graph} shows the obtained evolution graph, projected onto the 2-dimensional space defined by \textit{NetIncome} and \textit{TaxRate}, two of the descriptive variables in the dataset \texttt{EC}.
The series of the \textit{NetIncome} on the $ \mathcal{C}_2 \longrightarrow \mathcal{C}_6 \longrightarrow \mathcal{C}_9$ evolution path is $-0.09 \longrightarrow -0.04 \longrightarrow 0.15$, whereas the \textit{TaxRate} on the same evolution path is $0.08 \longrightarrow -0.04 \longrightarrow -0.06$.
It depicts the ``tax optimization'' undertook by most companies in the dataset: most companies arrive to decrease significantly their tax rate, while increasing the net income.
Phase $\mathcal{C}_3$ is a crisis phase, out of the 15 companies that enter it and only one exits it.
This seems to indicate that in the economical climate preceding the crisis of 2009, one of the ways to keep a company profitable was fiscal optimization.

\begin{table}[b]
\caption{Most common evolution phases in \texttt{EC}, described over the 7 dimensions of the dataset. The evolution path $ \mathcal{C}_2 \longrightarrow \mathcal{C}_6 \longrightarrow \mathcal{C}_9$ is the path followed by most companies}
\centering
\tabcolsep=0.11cm
\begin{tabular}{rrrrrrrrr}
  \toprule
  Ph. & Time & FCFF & TotalDebt & Revenues & NetCapExp & EBITDA & TaxRate & NetIncome \\ 
  \midrule
  $\mathcal{C}_2$ & 01/2003 & -0.00 & -0.01 & -0.02 & -0.00 & -0.04 & 0.08 & -0.09 \\ 
  $\mathcal{C}_3$ & 08/2003 & -0.94 & -0.06 & -1.82 & -0.67 & -2.02 & -0.07 & -4.04 \\ 
  $\mathcal{C}_6$ & 06/2004 & -0.01 & -0.01 & -0.02 & -0.04 & -0.02 & -0.04 & -0.04 \\ 
  $\mathcal{C}_9$ & 06/2006 & 0.05 & 0.01 & 0.07 & 0.04 & 0.07 & -0.06 & 0.15 \\ 
   \bottomrule
\end{tabular}
\label{tab:centroids-EC}
\end{table}

\subsection{Quantitative Results}
\label{subsec:quantitative-results}

To handle the initialization bias present in clustering, we have constructed 20 sets of initial \rev{prototypes}.
Each of the tested algorithms was initialized identically, with each of the sets of initial \rev{prototypes} and only the average values of the measures are reported.
The performances of six algorithms are compared:
\squishlisttwo 
	\item \textbf{Simple \KMeans{}}~\cite{MAC67} clusters the observations based solely on their resemblance in the multidimensional space.
	Optimizes \textit{MDvar};
	
	\item \textbf{Temporal-Driven \KMeans{}}~\cite{RIZ12} uses \KMeans{} with the the temporal-aware measure.
	Optimizes \textit{MDvar} and \textit{Tvar}.
	Parameters: $\alpha=0$ and $\beta=0$;
	
	\item \textbf{Constrained \KMeans{}}~\cite{RIZ12} uses the Euclidean distance and a penalty. 
	Optimizes \textit{MDvar} and \textit{ShaP}.
	Parameters: $\alpha=1$, $\beta = 0.0005$ and $\delta = 3$; 
	
	\item \textbf{tcK-Means}~\cite{LIN06} is a temporal constrained clustering algorithm.
	It uses a threshold penalty function, adapted to the multi-entity case.
	Optimizes \textit{\mbox{MDvar}} and \textit{ShaP}.
	Parameters: $\alpha^* = 2, d^* = 4$.
	
	\item \textbf{\TDCKMeans{}}~\cite{RIZ12} uses the temporal-aware dissimilarity measure, as well as contiguity constraints.
	Optimizes \textit{MDvar}, \textit{Tvar} and \textit{ShaP}.
	Parameters: $\alpha=0.95$, $\beta = 0.0002$ and $\delta = 3$;
	
	\item \textbf{\cluspath{}} is the algorithm we propose in Sect.~\ref{sec:our-proposal}.
	Unlike the aforementioned algorithms, \cluspath{} is the sole algorithm to infer a graph structure for the clusters during the clustering.
	Optimizes \textit{MDvar}, \textit{Tvar}, \textit{ShaP} and \textit{SPass}.
	Parameters: determined automatically using the heuristic in Sect.~\ref{subsec:choose-params-evolutionary}.
\squishend 

\begin{figure}[tbp]
	\centering

	\subfloat[] {
		\includegraphics[width=0.31\textwidth]{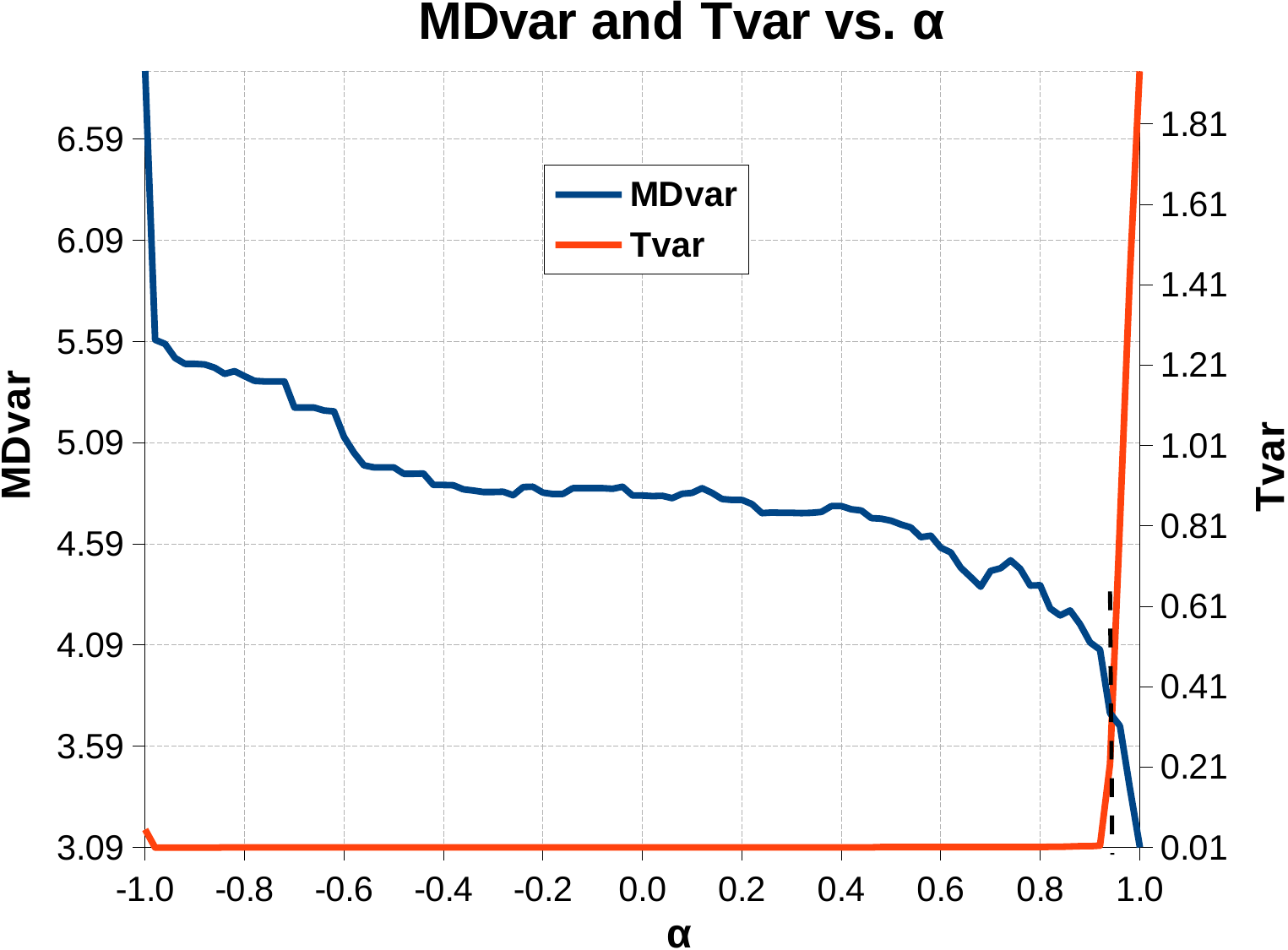}%
		\label{subfig:ec-determine-alpha}
	}
	\hfill
	\subfloat[]{
		\includegraphics[width=0.31\textwidth]{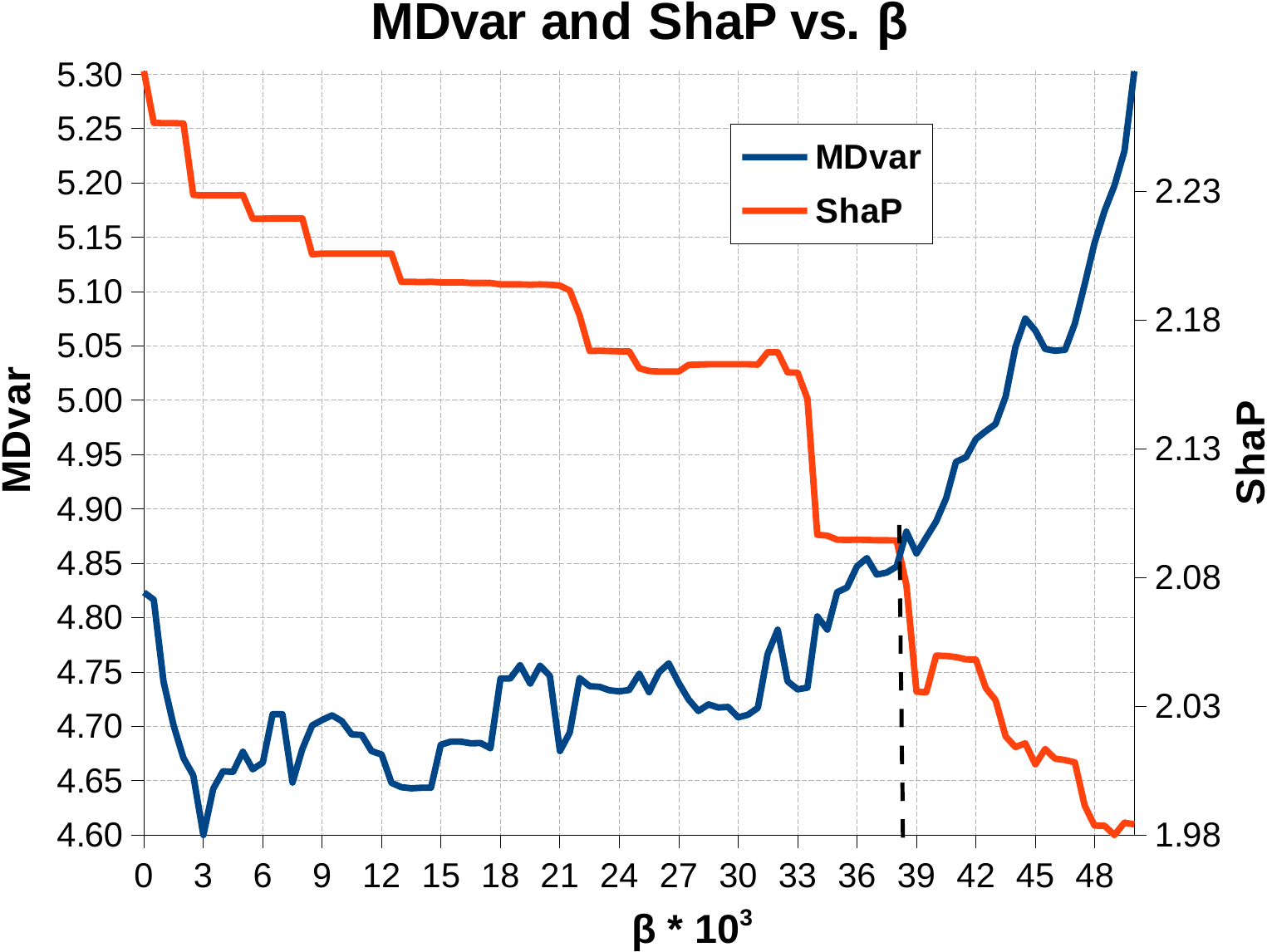}%
		\label{subfig:ec-determine-beta}
	}
	\hfill
	\subfloat[] {
		\includegraphics[width=0.31\textwidth]{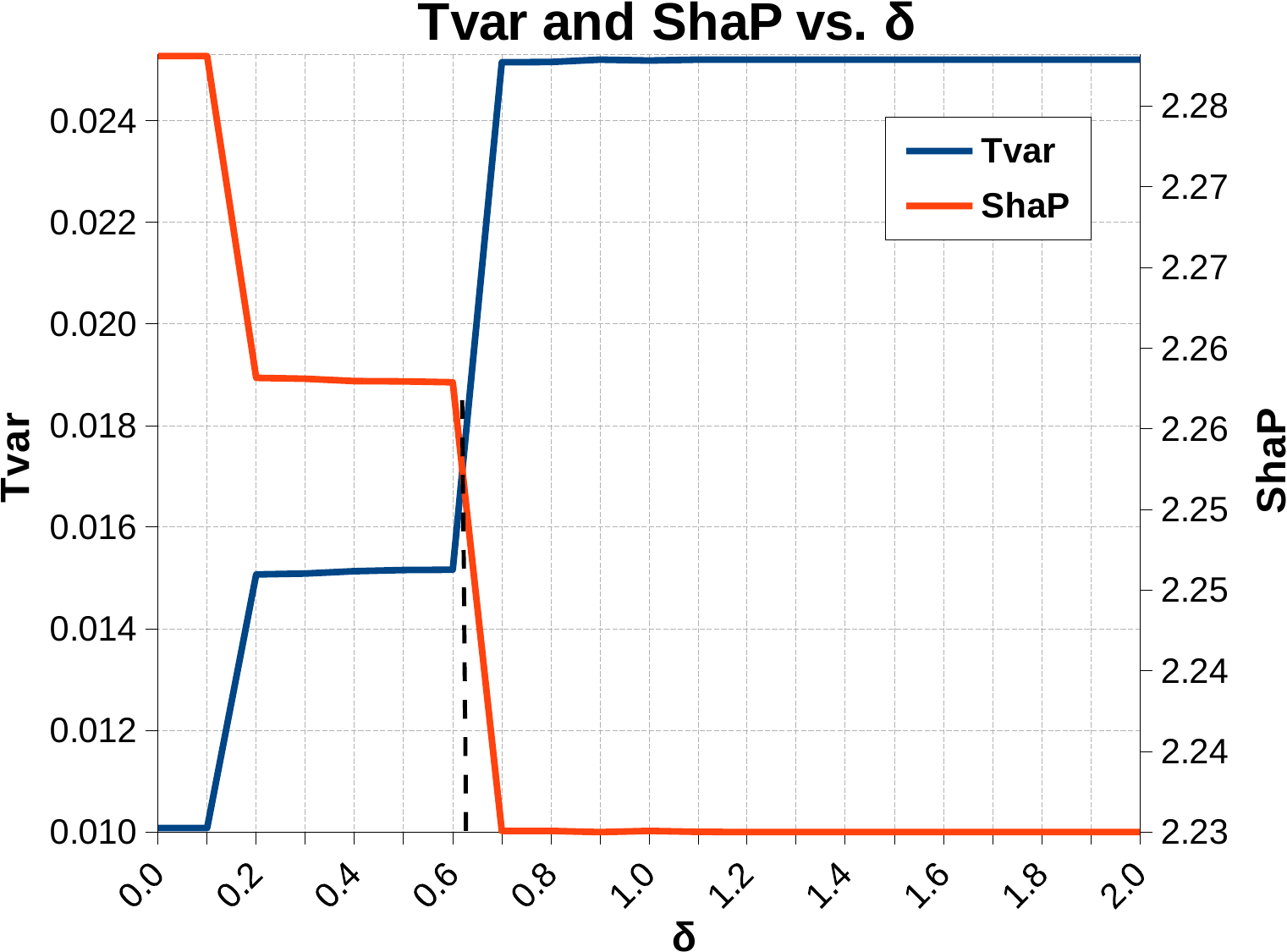}%
		\label{subfig:ec-determine-delta}
	}

	\caption{Determine, using the heuristics described in the original paper, the values of \TDCKMeans{}' parameters for dataset \texttt{EC}: $\alpha$ (a), $\beta$ (b) and $\delta$ (c)}
	\label{fig:ec-params-tdckmeans}
\end{figure}

\begin{table}[bp]
\caption{Mean value and standard deviation of evaluation measures for the different algorithms. All measures need to be minimized (best results in bold)}
\centering
\small
\tabcolsep=0.11cm
\begin{tabular}{llrr|rr|rr|rr}
	\toprule
	& \multicolumn{1}{c}{Algorithm} & \multicolumn{2}{c}{\textit{MDvar}} & \multicolumn{2}{c}{\textit{Tvar}} & \multicolumn{2}{c}{\textit{ShaP}} & \multicolumn{2}{c}{\textit{SPass}} \\ 
 \midrule

\multirow{6}{*}{\begin{sideways}\texttt{CPDS1}\end{sideways}} & \KMeans{} & \textbf{114.89} & \textit{3.8} & 46.08 & \textit{8.03} & 1.80 & \textit{0.18} & 3.19 & \textit{0.63} \\ 

& Temporal Driven K-Means & 125.32 & \textit{3.4} & \textbf{4.56} & \textit{0.35} & 2.96 & \textit{0.07} & 1.52 & \textit{0.31} \\ 

& Constrained K-Means & 140.26 & \textit{17.51} & 163.60 & \textit{31.19} & \textbf{0.51} & \textit{0.34} & 1.13 & \textit{0.49} \\

& tcK-Means & 131.54 & \textit{10.8} & 156.15 & \textit{19.23} & 0.61 & \textit{0.20} & 2.15 & \textit{0.51} \\  
 
& \TDCKMeans{} & 121.27 & \textit{4.34} & 38.58 & \textit{7.16} & 1.60 & \textit{0.13} & 1.52 & \textit{0.55} \\ 

& \cluspath{} & 118.68 & \textit{5.18} & 6.26 & \textit{3.21} & 2.81 & \textit{0.31} & \textbf{0.86} & \textit{0.23} \\ 

\midrule

\multirow{6}{*}{\begin{sideways}\texttt{EC}\end{sideways}} & \KMeans{} & \textbf{3.06} & \textit{0.28} & 1.95 & \textit{0.01} & 0.15 & \textit{0.15} & 8.41 & \textit{1.50} \\ 

& Temporal Driven K-Means & 4.83 & \textit{0.43} & \textbf{0.01} & \textit{0.03} & 2.28 & \textit{0.12} & 45.99 & \textit{9.63} \\ 

& Constrained K-Means & 4.96 & \textit{0.69} & 1.99 & \textit{0.01} & 0.21 & \textit{0.03} & 4.49 & \textit{6.56} \\ 

& tcK-Means & 4.29 & \textit{0.34} & 1.98 & \textit{0.01} & \textbf{0.04} & \textit{0.01} & 20.09 & \textit{14.61} \\ 

& \TDCKMeans{} & 4.41 & \textit{0.55} & 0.07 & \textit{0.06} & 2.14 & \textit{0.17} & 5.03 & \textit{1.29} \\ 

& \cluspath{} & 3.85 & \textit{0.59} & 0.60 & \textit{0.25} & 0.97 & \textit{0.22} & \textbf{4.40} & \textit{1.25} \\ 

\bottomrule
\end{tabular}
\label{tab:comparison-algos}
\end{table}

\textbf{Obtained results.}
The parameters of all algorithms, except \cluspath{}, are determined as shown in their original articles.
For example, in Figure~\ref{fig:ec-params-tdckmeans} we reproduce the heuristic of determining \TDCKMeans{}' parameters on \texttt{EC}.
Table~\ref{tab:comparison-algos} shows the average values of the measures, as well as the standard deviation (in italic) obtained by each algorithm.
The best results on each measure are indicated in boldface.
Note that, while \cluspath{} is designed to provide a compromise between the learning objectives, Simple \KMeans{}, Temporal-Driven \KMeans{} and Constrained \KMeans{} are designed to optimize mainly one component.
Not surprisingly, they show the best scores for, respectively, \textit{MDvar}, \textit{Tvar} and \textit{ShaP}.
\cluspath{} shows the best \textit{SPass} score, proving that constructing the structure between clusters during the clustering process results in evolution paths with smoother passages.
Both \TDCKMeans{} and \cluspath{} seek to provide a trade-off between the measures, but \cluspath{} consistently outperforms \TDCKMeans{}, except for the temporal variance on \texttt{EC}.
Overall, \cluspath{} succeeds in providing a good trade-off between the different contradicting measures, obtaining the best \textit{SPass} value and limited loss on the other measures.

\vspace*{-0.03\textheight}
\subsection{Impact of Parameters and Result Stability}
\label{subsec:parameters-stability}
\vspace*{-0.01\textheight}

\textcolor{black}{We launch the evolutionary heuristic on \texttt{CPDS1} 100 times with the same parameters of the evolutionary algorithm and with the same initial \rev{prototypes} for each execution of \cluspath{}.
This allows to i) assess the correlations between the chosen parameters and the obtained measures and ii) study the stability of the chosen solution, while lowering th impact of initialization randomness present in \KMeans{}-like algorithms.
}

\textbf{Impact of parameters on the evaluation measures}
Table~\ref{tab:correlation-matrix} shows the Pearson correlation between i) the parameters and the measure, ii) between the parameters and iii) between the measures.
Statistically significant correlations (with $p = 0.05$) are shown in boldface.
The table on the right shows that the evaluation measures are, by pairs of two, correlated among themselves.
The pairs a) \textit{MDvar} and \textit{ShaP} and b) \textit{Tvar} and \textit{SPass} are correlated positively.
Conversely, \textit{MDvar} and \textit{ShaP} seem to be negatively correlated with {Tvar} and \textit{SPass}.
The result is consistent with the multiobjective optimization task:
all solutions are scattered closely around the optimum solution and improving a measure mechanically involves degrading others, hence the negative correlations.
\textcolor{black}{
The parameters with the most impact in \cluspath{} are $\alpha$ and $\beta$.
They also have a weak correlation among themselves.
$\alpha$ is statistically significantly correlated, negatively with \textit{MDvar} and \textit{ShaP}, and positively with \textit{Tvar}.
This was expected, considering that $\alpha$ is a slider variable, giving priority to the temporal component for higher values.
$\beta$ is negatively correlated with \textit{MDvar} and \textit{SPass} and positively correlated with \textit{ShaP}: the higher $\beta$, the higher the contiguity penalty, which results in a more contiguous segmentation of observations for each entity.
}

\begin{table}[bp]
\caption{Correlation matrix i) between parameters and quality measures and ii) between parameters (left table) and iii) between quality measures (right table).
}
\centering
\tabcolsep=0.07cm
\small
\begin{tabular}{l|rrrr|rrrrrr}
\toprule

\multicolumn{ 1}{c}{}& \multicolumn{1}{c}{\textit{MDvar}} & \multicolumn{1}{c}{\textit{Tvar}} & \multicolumn{1}{c}{\textit{ShaP}} & \multicolumn{1}{c}{\textit{SPass}} & \multicolumn{1}{c}{$\alpha$} & \multicolumn{1}{c}{$\beta$} & \multicolumn{1}{c}{$\delta$} & \multicolumn{1}{c}{$\lambda_1$} & \multicolumn{1}{c}{$\lambda_2$} & \multicolumn{1}{c}{$\lambda_3$} \\ 

\midrule

$\alpha$ & \textbf{-0.92} & \textbf{0.72} & \textbf{-0.90} & 0.14 & \textbf{1.00} & 0.34 & 0.15 & -0.07 & -0.08 & -0.08 \\ 

$\beta$ & \textbf{-0.31} & 0.09 & \textbf{0.30} & \textbf{-0.21} &  & \textbf{1.00} & -0.07 & -0.20 & -0.12 & -0.06 \\

$\delta$ & -0.17 & 0.05 & -0.15 & 0.14 &  &  & \textbf{1.00} & -0.08 & 0.07 & -0.11 \\ 

$\lambda_1$ & 0.05 & -0.06 & 0.08 & 0.13 &  &  &  & \textbf{1.00} & 0.09 & 0.03 \\

$\lambda_2$ & 0.11 & -0.01 & 0.09 & -0.09 &  &  &  &  & \textbf{1.00} & 0.00 \\ 

$\lambda_3$ & 0.03 & -0.06 & 0.03 & 0.07 &  &  &  &  &  & \textbf{1.00} \\ 

\bottomrule
\end{tabular}
\hfill
\begin{tabular}{lrrrr}
\toprule
 & \textit{MDvar} &  \textit{Tvar} & \textit{ShaP} & \textit{SPass} \\ 
\midrule
 
\textit{MDvar} & \textbf{1.00} & \textbf{-0.68} & \textbf{0.95} & \textbf{-0.31} \\ 
\textit{Tvar} &  & \textbf{1.00} & \textbf{-0.76} & \textbf{0.34} \\ 
\textit{ShaP} &  &  & \textbf{1.00} & \textbf{-0.37} \\ 
\textit{SPass} &  &  &  & \textbf{1.00} \\ 

\bottomrule
\end{tabular}
\label{tab:correlation-matrix}
\end{table}

\begin{figure}[tbp]
\centering
	\subfloat[] {
		\includegraphics[height=0.152\textheight]{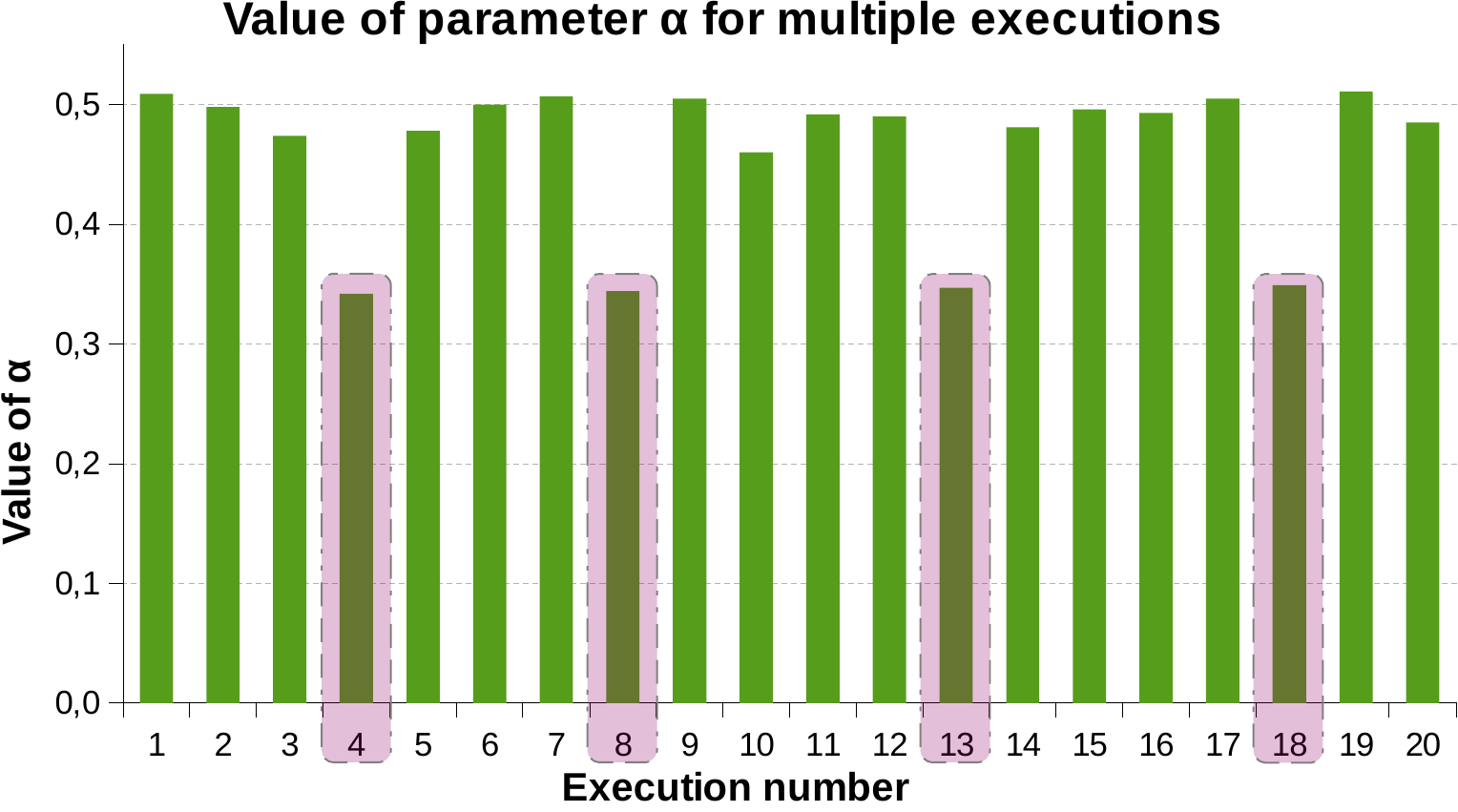}%
		\label{subfig:stability-study-Alpha}
	}
	\hfill
	\subfloat[]{
		\includegraphics[height=0.152\textheight]{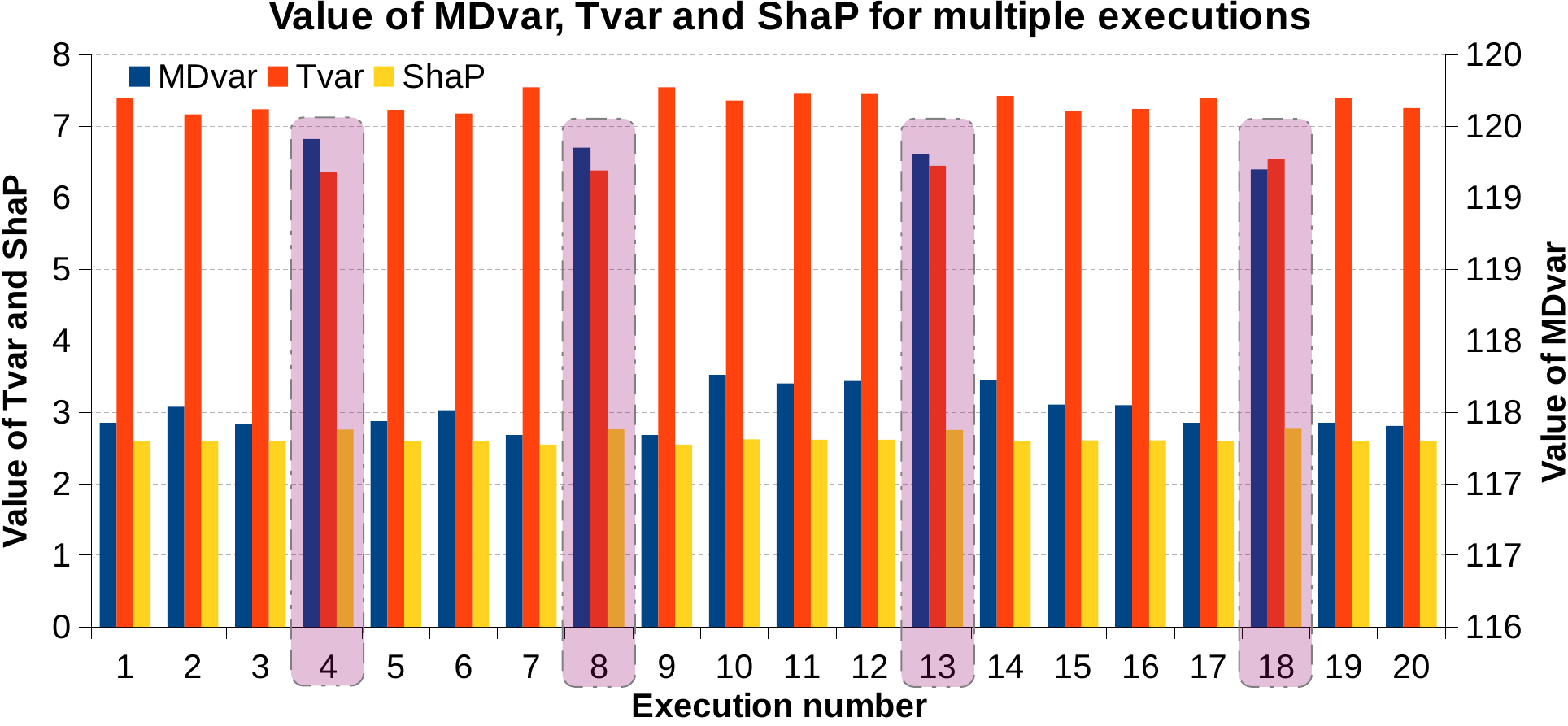}%
		\label{subfig:stability-study-measures}
	}

	\caption{Two local optima: the region identified by negative values of $\alpha$ is highlighted (a) and the corresponding values for \textit{MDvar}, \textit{Tvar} and \textit{ShaP} (b).}
	\label{fig:stability-study}
\end{figure}

\begin{figure}[htbp]
	\centering
	\includegraphics[angle=90,origin=c,height=0.8\textheight]{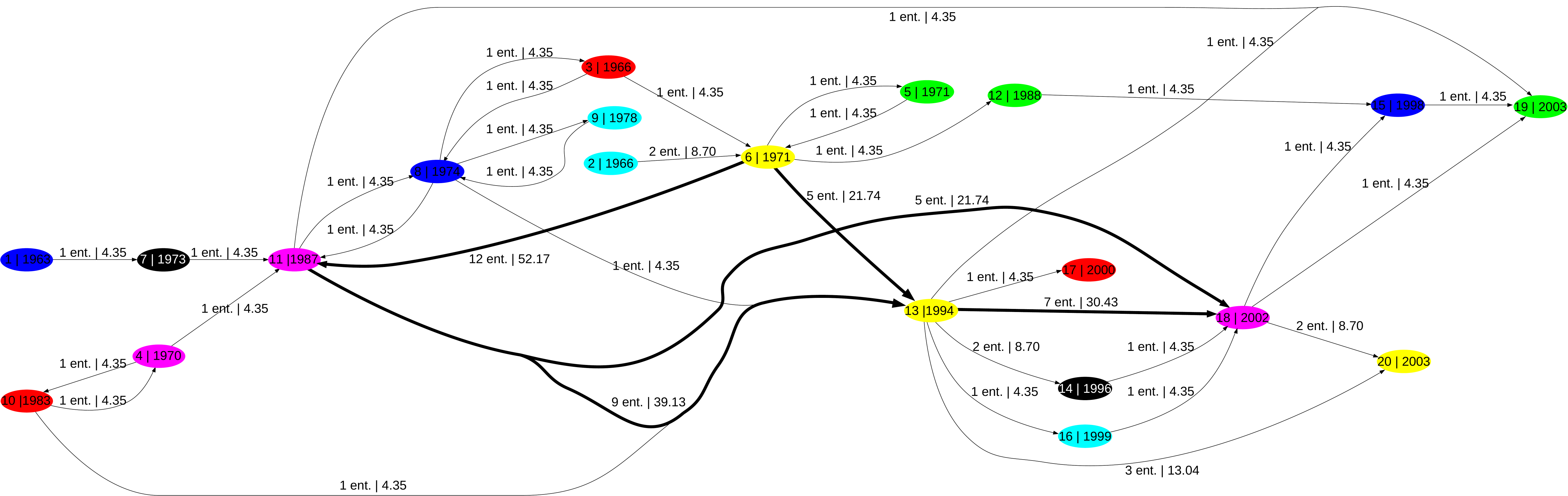}
	\\
	\caption{Graph structure constructed \textit{a posteriori} by \TDCKMeans{}, on \textit{Comparative Political Data Set I} with 20 clusters. }
	\label{fig:TDCKMeans-graph}
\end{figure}

\textbf{Stability of the chosen solution for \cluspath{}}
\rev{We assess how stable are the solutions constructed by the evolutionary technique by studying the variability of the obtained parameters of \cluspath{} and the the values of evaluation measures.
For parameter $\alpha$, for most of the 100 execution described in this subsection, its values are distributed uniformly around $0.48$.
The exception are four cases in which $\alpha$ takes values around $0.35$.
These four ``outlier'' values of $\alpha$, together with another 16 samples ``regular'' values (only 16 for readability purposes) are shown in Fig.~\ref{subfig:stability-study-Alpha}.}
There are no intermediary values between the two levels, which indicates that the Pareto front has two regions close to the ideal point: a larger one defined by values of $\alpha$  around $0.48$ and a second, considerably smaller one, defined by $\alpha \approx 0.35$ (highlighted in Fig.~\ref{subfig:stability-study-Alpha}). 
Fig.~\ref{subfig:stability-study-measures} highlights the corresponding values of \textit{MDvar}, \textit{Tvar} and \textit{ShaP}.
This local minimum region of the Pareto front presents consistently elevated values of \textit{MDvar}, slightly higher values of \textit{ShaP} and lower values of \textit{Tvar}.
These observations are consistent with the conclusions of the previous paragraph, considering that $\alpha$ is correlated, i) negatively with \textit{MDvar} and \textit{ShaP} and ii) positively with \textit{Tvar}.
The existence of two local optima is confirmed by Table~\ref{tab:removing-local-optimum}, \rev{
which shows the mean and the coefficient of variation for the six parameters and four measures of \cluspath{} over all the executions (denoted by \textbf{*}) and after removing the four solutions corresponding to the local minima (denoted by \textbf{**}).
The coefficients of variation of the evaluation measures for case \textbf{*} consistently decrease in the \textbf{**} case.
This proves that the removed solutions were caught in a local optimum region and the remaining solutions are grouped even more densely in the 4-dimensional space of the measures.
}
%

\begin{table}[tbp]
\caption{Result stability: averages and coefficients of variation for all solutions (rows with \textbf{*}) and after removing the local optimum corresponding to lower $\alpha$ values (rows with \textbf{**}) }
\small
\centering
\tabcolsep=0.11cm
\begin{tabular}{lrrrrrrr|rrrr}
\toprule
&  \multicolumn{1}{l}{} & \multicolumn{1}{c}{$\alpha$} & \multicolumn{1}{c}{$\beta \times 10^5 $} & \multicolumn{1}{c}{$\delta$} & \multicolumn{1}{c}{$\lambda_1$} & \multicolumn{1}{c}{$\lambda_2$} & \multicolumn{1}{c}{$\lambda_3$} & \multicolumn{ 1}{c}{MDvar} & \multicolumn{ 1}{c}{Tvar} & \multicolumn{ 1}{c}{ShaP} & \multicolumn{ 1}{c}{SPass} \\  
 
\midrule
\multirow{2}{*}{\begin{sideways}\textbf{*}\end{sideways}}  & Average & 0.477 & 7.727 & 3.02 & 484.74 & 536.51 & 522.35 & 117.69 & 7.33 & 2.61 & 0.76 \\ 			
 
& Coef. var. & 7.11\% & 54.48\% & 18.35\% & 55.97\% & 52.76\% & 56.34\% & 0.32\% & 3.12\% & 1.35\% & 1.83\% \\ 

\midrule
\multirow{2}{*}{\begin{sideways}\textbf{**}\end{sideways}} & Average & 0.480 & 7.5 & 2.99 & 587.54 & 535.91 & 502.65 & 117.65 & 7.39 & 2.61 & 0.77 \\
							
& Coef. var. & 4.28\% & 59.63\% & 16.24\% & 37.22\% & 34.34\% & 61.48\% & \textbf{0.17\%} & \textbf{2.07\%} & \textbf{0.85\%} & \textbf{0.91\%} \\ 

\bottomrule
\end{tabular}
\label{tab:removing-local-optimum}
\end{table}

\section{Conclusion and Future Work}
\label{sec:conclusion}
\vspace*{-0.01\textheight}

In this paper, we have studied the construction of typical evolution paths followed by a collection of entities.
We have proposed a novel algorithm, \cluspath{}, that partitions the observations belonging to entities into clusters, coherent in both the descriptive and temporal spaces.
The connexions between clusters are inferred during the clustering process and successions of linked clusters are interpreted as evolution paths.
A semi-supervised technique is used to leverage the strength of the links between clusters in the assignment of observations.
An evolutionary technique is used to find the set of optimum parameters and choose the ``best'' trade-off of measures.
We perform experiments on two real-live datasets, one issued from political sciences and the other issued from economics.
We have shown how complex notions, such as socio-economical models (\textit{i.e.}, the ``Swedish'' model) or tax policies (\textit{i.e.}, the tax optimization performed by companies) can be detected from the temporal evolution of descriptive features.

\textcolor{black}{
The main novelty of \cluspath{} over other approaches (such as co-clustering) is that i) it joins the temporal and descriptive features in the same objective function and ii) it combines into the same optimization procedure the descriptive-temporal construction of the \rev{prototypes} with the inference of the relations between clusters.
The major advantage over constructing each component sequentially (like in \TDCKMeans{} with \textit{a posteriori} graph structure construction) is that the content of a cluster and relations between clusters influence each other during the optimization process.
\cluspath{} is based on a ``slow changing world'' hypothesis, which assumes that changes in the population are gradual and ``smooth''.
This hypothesis holds for many application domains, \textit{e.g.}, scientific discussion topics, online communities \textit{etc.}
In applications in which this hypothesis does not hold (\textit{e.g.}, stock market transactions, in which it is desirable to detect sudden changes), \cluspath{} can still be used, by lowering the degree in which this hypothesis is enforced.
}

\textbf{Future work.} 
We are currently experimenting with applying the algorithm to other applications, \textit{e.g.}, detection of social roles in social networks, by passing through temporal behavioral roles.
A social role is defined as a typical succession of behavioral roles.
Another direction of research is describing the clusters with an easily comprehensible description by introducing temporal information into an unsupervised feature construction algorithm.
\textcolor{black}{
Finally, it would be useful to compare the solution constructed by \cluspath{} with those issued by algorithms for detecting trajectories of moving clusters (\textit{e.g.},~\cite{Kalnis2005}).
}

\vspace*{-0.025\textheight}
\section{Compliance with Ethical Standards}
\vspace*{-0.01\textheight}
\textbf{Conflict of Interest:} The authors declare that they have no conflict of interest. \\
\textbf{Research involving Human Participants and/or Animals:} The authors declare that no part of the research presented in this manuscript involved any humans or animals.

\begin{acknowledgements}
NICTA is funded by the Australian Government through the Department of Communications and the Australian Research Council through the ICT Centre of Excellence Program.
\end{acknowledgements}
 
{ \scriptsize
\begin{thebibliography}{26}
\providecommand{\natexlab}[1]{#1}
\providecommand{\url}[1]{{#1}}
\providecommand{\urlprefix}{URL }
\expandafter\ifx\csname urlstyle\endcsname\relax
  \providecommand{\doi}[1]{DOI~\discretionary{}{}{}#1}\else
  \providecommand{\doi}{DOI~\discretionary{}{}{}\begingroup
  \urlstyle{rm}\Url}\fi
\providecommand{\eprint}[2][]{\url{#2}}

\bibitem[{Araujo and Kamel(2014)}]{Araujo2014}
Araujo R, Kamel MS (2014) {Semi-supervised Kernel-Based Temporal Clustering}.
  In: International Conference on Machine Learning and Applications, IEEE,
  ICMLA~'14, pp 123--128

\bibitem[{{Armingeon, Klaus, Christian Isler, Laura Kn{\"{o}}pfel} and
  Engler(2011)}]{ARM11}
{Armingeon, Klaus, Christian Isler, Laura Kn{\"{o}}pfel} DW, Engler S (2011)
  {Comparative Political Data Set 1960-2009}. University of Berne.

\bibitem[{Chakrabarti et~al(2006)Chakrabarti, Kumar, and Tomkins}]{CHA06a}
Chakrabarti D, Kumar R, Tomkins A (2006) {Evolutionary Clustering}. In:
  International Conference on Knowledge Discovery and Data Mining, ACM, SIGKDD
  '06, pp 554--560

\bibitem[{Chi et~al(2007)Chi, Song, Zhou, Hino, and Tseng}]{CHI07}
Chi Y, Song X, Zhou D, Hino K, Tseng BL (2007) {Evolutionary Spectral
  Clustering by Incorporating Temporal Smoothness}. In: International
  Conference on Knowledge Discovery and Data Mining (KDD), San Jose, USA, pp
  153--162

\bibitem[{{De la Torre} and Agell(2007)}]{TOR07}
{De la Torre} F, Agell C (2007) {Multimodal Diaries}. In: Multimedia and Expo,
  IEEE, pp 839--842

\bibitem[{{De Smet} and Eppe(2009)}]{SME09}
{De Smet} Y, Eppe S (2009) {Multicriteria Relational Clustering: The Case of
  Binary Outranking Matrices}. In: Evolutionary Multi-Criterion Optimization,
  vol 5467, pp 380--392

\bibitem[{Deb et~al(2002)Deb, Pratap, Agarwal, and Meyarivan}]{DEB02}
Deb K, Pratap A, Agarwal S, Meyarivan T (2002) {A fast and elitist
  multiobjective genetic algorithm: NSGA-II}. Evolutionary Computation
  6(2):182--197

\bibitem[{Dunn(1973)}]{DUN73}
Dunn JC (1973) {A Fuzzy Relative of the ISODATA Process and Its Use in
  Detecting Compact Well-Separated Clusters}. Journal of Cybernetics
  3(3):32--57

\bibitem[{Erixon(2000)}]{Erixon2000}
Erixon L (2000) {A Swedish Economic Policy: The Theory, Application and
  Validity of the Rehn-Meidner Model}. Tech. rep., Department of Economics,
  Stockholm University

\bibitem[{Gaffney and Smyth(1999)}]{Gaffney1999}
Gaffney S, Smyth P (1999) {Trajectory clustering with mixtures of regression
  models}. In: International Conference on Knowledge Discovery and Data Mining,
  ACM Press, New York, USA, SIGKDD~'99, pp 63--72, \doi{10.1145/312129.312198}

\bibitem[{Halsall-Whitney and Thibault(2006)}]{HAL06}
Halsall-Whitney H, Thibault J (2006) {Multi-objective optimization for chemical
  processes and controller design : Approximating and classifying the pareto
  domain}. Computers \& Chemical Engineering 30(6-7):1155--1168

\bibitem[{Kafafy et~al(2011)Kafafy, Bounekkar, and Bonnevay}]{KAF11}
Kafafy A, Bounekkar A, Bonnevay S (2011) {A hybrid evolutionary metaheuristics
  (HEMH) applied on 0/1 multiobjective knapsack problems}. In: Genetic and
  Evolutionary Computation, ACM Press, New York, USA, GECCO~'11, p 497

\bibitem[{Kalnis et~al(2005)Kalnis, Mamoulis, and Bakiras}]{Kalnis2005}
Kalnis P, Mamoulis N, Bakiras S (2005) {On Discovering Moving Clusters in
  Spatio-temporal Data}. In: {Bauzer Medeiros} C, Egenhofer M, Bertino E (eds)
  Advances in Spatial and Temporal Databases, Lecture Notes in Computer
  Science, vol 3633, Springer Berlin Heidelberg, chap~21, pp 364--381

\bibitem[{Liang et~al(2013)Liang, Tomioka, Murata, Asaoka, and
  Yamanishi}]{Liang2013}
Liang Z, Tomioka R, Murata H, Asaoka R, Yamanishi K (2013) {Quantitative
  Prediction of Glaucomatous Visual Field Loss from Few Measurements}. In:
  International Conference on Data Mining, ICDM~'13, pp 1121--1126

\bibitem[{Lin and Hauptmann(2006)}]{LIN06}
Lin WH, Hauptmann A (2006) {Structuring continuous video recordings of everyday
  life using time-constrained clustering}. In: Chang EY, Hanjalic A, Sebe N
  (eds) Multimedia Content Analysis, Management, and Retrieval, pp
  60,730D--60,730D--9

\bibitem[{MacQueen(1967)}]{MAC67}
MacQueen J (1967) {Some methods for classification and analysis of multivariate
  observations}. In: Berkeley Symposium on Mathematical Statistics and
  Probability, vol~1, pp 281--297

\bibitem[{Mih\u{a}i\c{t}\u{a} et~al(2014)Mih\u{a}i\c{t}\u{a}, Camargo, and
  Lhoste}]{MIH14}
Mih\u{a}i\c{t}\u{a} AS, Camargo M, Lhoste P (2014) {Optimization of a complex
  urban intersection using discrete event simulation and evolutionary
  algorithms}. In: International Federation of Automatic Control, IFAC'14,
  vol~19, pp 8768--8774

\bibitem[{Rizoiu et~al(2012)Rizoiu, Velcin, and Lallich}]{RIZ12}
Rizoiu MA, Velcin J, Lallich S (2012) {Structuring typical evolutions using
  Temporal-Driven Constrained Clustering}. In: International Conference on
  Tools with Artificial Intelligence, IEEE, Athens, Greece, ICTAI~'12, vol~1,
  pp 610--617

\bibitem[{Rizoiu et~al(2014)Rizoiu, Velcin, and Lallich}]{Rizoiu2014b}
Rizoiu MA, Velcin J, Lallich S (2014) {How to Use Temporal-Driven Constrained
  Clustering to Detect Typical Evolutions}. International Journal on Artificial
  Intelligence Tools 23(04):1460,013

\bibitem[{Rizoiu et~al(2016)Rizoiu, Velcin, Bonnevay, and
  Lallich}]{supplemental}
Rizoiu MA, Velcin J, Bonnevay S, Lallich S (2016) Supplementary material: A
  temporal-driven clustering solution to inferring typical evolution paths.
  \url{http://goo.gl/KCWrSM}

\bibitem[{Rocha et~al(2013)Rocha, Dias, and Dimas}]{ROC13}
Rocha C, Dias LC, Dimas I (2013) {Multicriteria Classification with Unknown
  Categories: A Clustering-Sorting Approach and an Application to Conflict
  Management}. Journal of Multi-Criteria Decision Analysis 20(1-2):13--27

\bibitem[{Sawaragi et~al(1985)Sawaragi, Nakayama, and Tanino}]{SAW85}
Sawaragi Y, Nakayama H, Tanino T (1985) {Theory of multiobjective
  optimization}, vol 176. Academic Press New York

\bibitem[{Siddiqui et~al(2012)Siddiqui, Oliveira, Gama, and
  Spiliopoulou}]{Siddiqui2012}
Siddiqui ZF, Oliveira M, Gama J, Spiliopoulou M (2012) {Where Are We Going?
  Predicting the Evolution of Individuals}. In: Hollm{\'{e}}n J, Klawonn F,
  Tucker A (eds) Advances in Intelligent Data Analysis V, Lecture Notes in
  Computer Science, vol 7619, Springer Berlin Heidelberg, pp 357--368

\bibitem[{Wagstaff et~al(2001)Wagstaff, Cardie, Rogers, and Schroedl}]{WAG01}
Wagstaff K, Cardie C, Rogers S, Schroedl S (2001) {Constrained K-means
  Clustering with Background Knowledge}. In: International Conference on
  Machine Learning, ICML~'01, pp 577--584

\bibitem[{Xu et~al(2012)Xu, Zhang, Yu, and Long}]{XU12}
Xu T, Zhang Z, Yu PS, Long B (2012) {Generative models for evolutionary
  clustering}. ACM Transactions on Knowledge Discovery from Data (TKDD) 6(2):7

\bibitem[{Zitzler et~al(2001)Zitzler, Laumanns, and Thiele}]{ZIT01}
Zitzler E, Laumanns M, Thiele L (2001) {SPEA2: Improving the strength Pareto
  evolutionary algorithm}. In: Evolutionary Methods for Design, Optimisation
  and Control with Applications to Industrial Problems, EUROGEN~'01, pp 95--100

\end{thebibliography}
}



\section*{Supplementary material}

# *Supplementary Material*: ClusPath: A Temporal-driven Clustering Solution to Inferring Typical Evolution Paths

**Marian-Andrei Rizoiu** [1,*], **Julien Velcin** [2], **Stéphane Bonnevay** [3] **and Stéphane Lallich** [2]

[1] *NICTA Research Lab & Australian National University, Canberra, Australia*
[2] *Laboratory ERIC, University Lumière Lyon 2, Lyon, France*
[3] *Laboratory ERIC, University Claude Bernard Lyon 1, Lyon, France*

Correspondence*:
Marian-Andrei Rizoiu
NICTA Research Lab, 7 London Circuit, Canberra, Australia,
Marian-Andrei.Rizoiu@nicta.com.au

The purpose of this document is to detail information, which would be i) cumbersome for the main article and ii) of limited interest for the main scientific message presented in the main article. This document does not introduce new notions and the information presented here-after is not necessary for the comprehension of the main article. We present it for completeness and reproducibility reasons.

We produce hereafter, the complete calculations of the prototypes update formulas, as well as the formulas for updating the adjacency matrix and runtimes for the algorithms.

## 1 INFERRING THE PROTOTYPES UPDATE FORMULAS

The objective function $\mathcal{J}$ that needs to be minimized using a gradient descend K-Means-like framework is (defined in Equation **8** in the Main Text (MT)):

$$\mathcal{J} = \lambda_1 T_1 + \lambda_2 T_2 + \lambda_3 T_3 =$$

$$= \lambda_1 \sum_{\mu_p \in \mathcal{M}} \sum_{x_i \in \mathcal{C}_p} \left( ||x_i - \mu_p||_{TA} + \sum_{\substack{x_k \in \mathcal{C}_q \\ q \neq p \\ x_i^\phi = x_k^\phi}}^{x_i^t < x_k^t} \beta * e^{-\frac{1}{2}\left(\frac{||x_i^t - x_k^t||}{\delta}\right)^2} \left(1 - a_{p,q}^2\right) \right) +$$

$$+ \lambda_2 \sum_{\substack{\mu_p \in \mathcal{M} \\ p \neq q}} \sum_{\mu_q \in \mathcal{M}} a_{p,q}^2 ||\mu_p - \mu_q||_{TA} +$$

$$+ \lambda_3 \sum_{\substack{\mu_p \in \mathcal{M} \\ p \neq q}} \sum_{\mu_q \in \mathcal{M}} a_{p,q}^2 inter_\phi^2(\mathcal{C}_p, \mathcal{C}_q) \ , \tag{1}$$

and the temporal-aware dissimilarity measure is defined as (in Equation 1 in MT):

$$||x_i - x_j||_{TA} = 1 - \left(1 - \gamma_d \frac{||x_i^d - x_j^d||^2}{\Delta d_{max}^2}\right)\left(1 - \gamma_t \frac{||x_i^t - x_j^t||^2}{\Delta t_{max}^2}\right) \ , \quad (2)$$

where the weight of the descriptive component and, respectively, the temporal component, are controlled by the parameter $\alpha \in [-1, 1]$:

$$\gamma_d = \begin{cases} 1 + \alpha, & \text{if } \alpha \in [-1, 0] \\ 1, & \text{if } \alpha \in (0, 1] \end{cases} \ ; \ \gamma_t = \begin{cases} 1, & \text{if } \alpha \in [-1, 0] \\ 1 - \alpha, & \text{if } \alpha \in (0, 1] \end{cases} .$$

As stated in the main article, calculating the update formulas for the prototypes boils down to recomputing each of the descriptive components ($\mu_j^d$) and temporal component ($\mu_j^t$) of the prototypes. More precisely, we are searching for the fixed point, by calculating the derivative of the function $\mathcal{J}$ of the variables $\mu_j^d$ and $\mu_j^t$. Therefore, the following system of equations needs to be solved:

$$\frac{\partial \mathcal{J}}{\partial \mu_j^d} = 0 \ ; \ \frac{\partial \mathcal{J}}{\partial \mu_j^t} = 0$$

We exemplify the calculation of the derivative of $\mathcal{J}$ of the variable $\mu_j^d$. From Equation 1 we obtain:

$$\frac{\partial \mathcal{J}}{\partial \mu_j^d} = \lambda_1 \frac{\partial T_1}{\partial \mu_j^d} + \lambda_2 \frac{\partial T_2}{\partial \mu_j^d} + \lambda_3 \frac{\partial T_3}{\partial \mu_j^d} \quad (3)$$

Knowing that the $inter_\phi$ function from the term $T_3$ is defined as:

$$inter_\phi(\mathcal{C}_p, \mathcal{C}_q) = 1 - \frac{|\{\phi_l \in \Phi | \mathcal{C}_p \xrightarrow{\phi_l} \mathcal{C}_q\}|}{|\Phi|} \ ,$$

and, therefore, it is independent of $\mu_j^d$, the following holds:

$$\frac{\partial T_3}{\partial \mu_j^d} = 0, \forall \mu_j^d \quad (4)$$

We augment the definition of the first two terms, by introducing the temporal-aware dissimilarity measure (in Equation 2) into the formula of the objective function (Equation 1):

$$T_1 = \sum_{\mu_p \in \mathcal{M}} \sum_{x_i \in \mathcal{C}_p} \left(1 - \left(1 - \gamma_d \frac{||x_i^d - \mu_p^d||^2}{\Delta d_{max}^2}\right)\left(1 - \gamma_t \frac{||x_i^t - \mu_p^t||^2}{\Delta t_{max}^2}\right) + \sum_{\substack{x_k \in \mathcal{C}_q \\ q \neq p \\ x_i^\phi = x_k^\phi}}^{x_i^t < x_k^t} \beta * e^{-\frac{1}{2}\left(\frac{||x_i^t - x_k^t||}{\delta}\right)^2}(1 - a_{p,q}^2)\right)$$

$$T_2 = \sum_{\mu_p \in \mathcal{M}} \sum_{\substack{\mu_q \in \mathcal{M} \\ p \neq q}} a_{p,q}^2 \left(1 - \left(1 - \gamma_d \frac{||\mu_p^d - \mu_q^d||^2}{\Delta d_{max}^2}\right)\left(1 - \gamma_t \frac{||\mu_p^t - \mu_q^t||^2}{\Delta t_{max}^2}\right)\right) \ .$$

We calculate the derivative of $T_1$. We observe that the penalty term $\sum_{\substack{x_i^t < x_k^t \\ x_k \in \mathcal{C}_q \\ q \neq p \\ x_i^\phi = x_k^\phi}} \beta * e^{-\frac{1}{2}\left(\frac{||x_i^t - x_k^t||}{\delta}\right)^2} \left(1 - a_{p,q}^2\right)$ is independent of $\mu_j^d$ and, therefore, it is nullified when calculating the derivative:

$$\frac{\partial T_1}{\partial \mu_j^d} = \frac{\partial}{\partial \mu_j^d} \left( \sum_{\mu_p \in \mathcal{M}} \sum_{x_i \in \mathcal{C}_p} \left(1 - \left(1 - \gamma_d \frac{||x_i^d - \mu_p^d||^2}{\Delta d_{max}^2}\right)\left(1 - \gamma_t \frac{||x_i^t - \mu_p^t||^2}{\Delta t_{max}^2}\right)\right)\right) \quad (5)$$

Considering that the double sum iterates over all the observations:

$$\sum_{\mu_p \in \mathcal{M}} \sum_{x_i \in \mathcal{C}_p} 1 = |\mathcal{X}|$$

Equation 5 becomes:

$$\frac{\partial T_1}{\partial \mu_j^d} = \frac{\partial |\mathcal{X}|}{\partial \mu_j^d} - \frac{\partial}{\partial \mu_j^d} \left( \sum_{\mu_p \in \mathcal{M}} \sum_{x_i \in \mathcal{C}_p} \left(1 - \gamma_d \frac{||x_i^d - \mu_p^d||^2}{\Delta d_{max}^2}\right)\left(1 - \gamma_t \frac{||x_i^t - \mu_p^t||^2}{\Delta t_{max}^2}\right)\right) =$$

$$= -\frac{\partial}{\partial \mu_j^d} \left( \sum_{\mu_p \in \mathcal{M}} \sum_{x_i \in \mathcal{C}_p} \left(1 - \gamma_d \frac{||x_i^d - \mu_p^d||^2}{\Delta d_{max}^2}\right)\left(1 - \gamma_t \frac{||x_i^t - \mu_p^t||^2}{\Delta t_{max}^2}\right)\right) =$$

$$= -\sum_{x_i \in \mathcal{C}_j} \frac{\partial}{\partial \mu_j^d} \left(1 - \gamma_d \frac{||x_i^d - \mu_j^d||^2}{\Delta d_{max}^2}\right)\left(1 - \gamma_t \frac{||x_i^t - \mu_j^t||^2}{\Delta t_{max}^2}\right) =$$

$$= -\frac{2\gamma_d}{\Delta d_{max}^2} \sum_{x_i \in \mathcal{C}_j} \left(x_i^d - \mu_j^d\right)\left(1 - \gamma_t \frac{||x_i^t - \mu_j^t||^2}{\Delta t_{max}^2}\right) \quad . \quad (6)$$

To ease the calculation of the derivative of $T_2$, we make the following notation:

$$s_{pq} = \left(1 - \gamma_t \frac{||\mu_p^t - \mu_q^t||^2}{\Delta t_{max}^2}\right) \quad . \quad (7)$$

Therefore:

$$\frac{\partial T_2}{\partial \mu_j^d} = \frac{\partial}{\partial \mu_j^d} \sum_{\mu_p \in \mathcal{M}} \sum_{\substack{\mu_q \in \mathcal{M} \\ p \neq q}} a_{p,q}^2 \left(1 - \left(1 - \gamma_d \frac{||\mu_p^d - \mu_q^d||^2}{\Delta d_{max}^2}\right) s_{pq}\right) \quad . \quad (8)$$

The term $\mu_j^d$ can appear twice: i) once in the first sum $\sum_{\mu_p \in \mathcal{M}}$, for $p = j$ and ii) once in the second sum $\sum_{\substack{\mu_q \in \mathcal{M} \\ p \neq q}}$, for $q = j$. Equation 8 can be rewritten as:

$$\frac{\partial T_2}{\partial \mu_j^d} = \sum_{\substack{\mu_q \in \mathcal{M} \\ q \neq j}} \frac{\partial}{\partial \mu_j^d} \left( a_{j,q}^2 \left(1 - \left(1 - \gamma_d \frac{||\mu_j^d - \mu_q^d||^2}{\Delta d_{max}^2}\right) s_{jq}\right)\right)$$

$$
+ \sum_{\substack{\mu_p \in \mathcal{M} \\ p \neq j}} \frac{\partial}{\partial \mu_j^d} \left( a_{p,j}^2 \left( 1 - \left( 1 - \gamma_d \frac{||\mu_p^d - \mu_j^d||^2}{\Delta d_{max}^2} \right) s_{pj} \right) \right) =
$$

$$
= \sum_{\substack{\mu_q \in \mathcal{M} \\ q \neq j}} \frac{\partial}{\partial \mu_j^d} \left( \frac{a_{j,q}^2 \gamma_d}{\Delta d_{max}^2} ||\mu_j^d - \mu_q^d||^2 s_{jq} \right) + \sum_{\substack{\mu_p \in \mathcal{M} \\ p \neq j}} \frac{\partial}{\partial \mu_j^d} \left( \frac{a_{p,j}^2 \gamma_d}{\Delta d_{max}^2} ||\mu_p^d - \mu_j^d||^2 s_{pj} \right) =
$$

$$
= \frac{2\gamma_d}{\Delta d_{max}^2} \left[ \sum_{\substack{\mu_q \in \mathcal{M} \\ q \neq j}} a_{j,q}^2 (\mu_j - \mu_q) s_{jq} - \sum_{\substack{\mu_p \in \mathcal{M} \\ p \neq j}} a_{p,j}^2 (\mu_p - \mu_j) s_{pj} \right] . \tag{9}
$$

We re-note the sum indexes so that $p = q$ and we observe that $s_{pq}$ defined in Equation 7 is symmetric: $s_{pq} = s_{qp}$. Consequently, we re-write Equation 9 as:

$$
\frac{\partial T_2}{\partial \mu_j^d} = \frac{2\gamma_d}{\Delta d_{max}^2} \left[ \mu_j^d \sum_{\substack{\mu_p \in \mathcal{M} \\ p \neq j}} a_{j,p}^2 s_{jp} - \sum_{\substack{\mu_p \in \mathcal{M} \\ p \neq j}} \mu_p^d a_{j,p}^2 s_{jp} + \mu_j^d \sum_{\substack{\mu_p \in \mathcal{M} \\ p \neq j}} a_{p,j}^2 s_{pj} - \sum_{\substack{\mu_p \in \mathcal{M} \\ p \neq j}} \mu_p^d a_{p,j}^2 s_{pj} \right] =
$$

$$
= \frac{2\gamma_d}{\Delta d_{max}^2} \left[ \mu_j^d \sum_{\substack{\mu_p \in \mathcal{M} \\ p \neq j}} s_{pj}(a_{j,p}^2 + a_{p,j}^2) - \sum_{\substack{\mu_p \in \mathcal{M} \\ p \neq j}} \mu_p^d s_{pj}(a_{j,p}^2 + a_{p,j}^2) \right] . \tag{10}
$$

By introducing Equations 4, 6 and 10 into Equation 3, we can calculate the formula of the derivative of function $\mathcal{J}$ of variable $\mu_j^d$ and we calculate the fixed point:

$$
\frac{\partial \mathcal{J}}{\partial \mu_j^d} = 0 \Leftrightarrow \frac{2\gamma_d}{\Delta d_{max}^2} \left[ \lambda_1 \mu_j^d \sum_{x_i \in \mathcal{C}_j} \left( 1 - \gamma_t \frac{||x_i^t - \mu_j^t||^2}{\Delta t_{max}^2} \right) - \lambda_1 \sum_{x_i \in \mathcal{C}_j} x_i^d \left( 1 - \gamma_t \frac{||x_i^t - \mu_j^t||^2}{\Delta t_{max}^2} \right) + \right.
$$

$$
\left. + \lambda_2 \mu_j^d \sum_{\substack{\mu_p \in \mathcal{M} \\ p \neq j}} s_{pj}(a_{j,p}^2 + a_{p,j}^2) - \lambda_2 \sum_{\substack{\mu_p \in \mathcal{M} \\ p \neq j}} \mu_p^d s_{pj}(a_{j,p}^2 + a_{p,j}^2) \right] = 0
$$

$$
\Leftrightarrow \mu_j^d \left[ \lambda_1 \sum_{x_i \in \mathcal{C}_j} \left( 1 - \gamma_t \frac{||x_i^t - \mu_j^t||^2}{\Delta t_{max}^2} \right) + \lambda_2 \sum_{\substack{\mu_p \in \mathcal{M} \\ p \neq j}} s_{pj}(a_{j,p}^2 + a_{p,j}^2) \right] =
$$

$$
= \lambda_1 \sum_{x_i \in \mathcal{C}_j} x_i^d \left( 1 - \gamma_t \frac{||x_i^t - \mu_j^t||^2}{\Delta t_{max}^2} \right) + \lambda_2 \sum_{\substack{\mu_p \in \mathcal{M} \\ p \neq j}} \mu_p^d s_{pj}(a_{j,p}^2 + a_{p,j}^2)
$$

$$\Leftrightarrow \mu_j^d = \frac{\lambda_1 \sum_{x_i \in \mathcal{C}_j} x_i^d \left(1 - \gamma_t \frac{||x_i^t - \mu_j^t||^2}{\Delta t_{max}^2}\right) + \lambda_2 \sum_{\substack{\mu_p \in \mathcal{M} \\ p \neq j}} \mu_p^d s_{pj}(a_{j,p}^2 + a_{p,j}^2)}{\lambda_1 \sum_{x_i \in \mathcal{C}_j} \left(1 - \gamma_t \frac{||x_i^t - \mu_j^t||^2}{\Delta t_{max}^2}\right) + \lambda_2 \sum_{\substack{\mu_p \in \mathcal{M} \\ p \neq j}} s_{pj}(a_{j,p}^2 + a_{p,j}^2)} \ . \quad (11)$$

By introducing the notation for $s_{pq}$ (defined in Equation 7) back into Equation 11, we obtain the final centroid update formula for the descriptive component:

$$\mu_j^d = \frac{\lambda_1 \sum_{x_i \in \mathcal{C}_j} x_i^d \left(1 - \gamma_t \frac{||x_i^t - \mu_j^t||^2}{\Delta t_{max}^2}\right) + \lambda_2 \sum_{\substack{\mu_p \in \mathcal{M} \\ p \neq j}} \mu_p^d \left(1 - \gamma_t \frac{||\mu_p^t - \mu_j^t||^2}{\Delta t_{max}^2}\right)(a_{j,p}^2 + a_{p,j}^2)}{\lambda_1 \sum_{x_i \in \mathcal{C}_j} \left(1 - \gamma_t \frac{||x_i^t - \mu_j^t||^2}{\Delta t_{max}^2}\right) + \lambda_2 \sum_{\substack{\mu_p \in \mathcal{M} \\ p \neq j}} \left(1 - \gamma_t \frac{||\mu_p^t - \mu_j^t||^2}{\Delta t_{max}^2}\right)(a_{j,p}^2 + a_{p,j}^2)} \ .$$

Similarly, we can deduce the update formula for the temporal component of prototypes:

$$\mu_j^t = \frac{\lambda_1 \sum_{x_i \in \mathcal{C}_j} x_i^t \left(1 - \gamma_d \frac{||x_i^d - \mu_j^d||^2}{\Delta d_{max}^2}\right) + \lambda_2 \sum_{\substack{\mu_p \in \mathcal{M} \\ p \neq j}} \mu_p^t \left(1 - \gamma_d \frac{||\mu_p^d - \mu_j^d||^2}{\Delta d_{max}^2}\right)(a_{j,p}^2 + a_{p,j}^2)}{\lambda_1 \sum_{x_i \in \mathcal{C}_j} \left(1 - \gamma_d \frac{||x_i^d - \mu_j^d||^2}{\Delta d_{max}^2}\right) + \lambda_2 \sum_{\substack{\mu_p \in \mathcal{M} \\ p \neq j}} \left(1 - \gamma_d \frac{||\mu_p^d - \mu_j^d||^2}{\Delta d_{max}^2}\right)(a_{j,p}^2 + a_{p,j}^2)} \ .$$

## 2 UPDATING THE ADJACENCY MATRIX

The objective function $\mathcal{J}$, in Equation 1, can be trivially minimized by setting $a_{p,q} = 0, \forall p, q \in [1, k]$. To avoid this situation, an additional constraint is imposed on the 1-norm of the adjacency matrix (in Equation 9 in MT):

$$||A||_1 = 1 \Leftrightarrow \sum_{p=1}^{k} \sum_{q=1}^{k} a_{p,q} = 1 \ . \quad (12)$$

The purpose of updating the adjacency matrix is to find the best adjacency matrix $A^*$ which minimizes the objective function $\mathcal{J}$:

$$A^* = \arg\min_A \mathcal{J} \ .$$

A classical strategy for finding the local maxima and minima of a function subject to equality constraints is method of Lagrange multipliers. This constructs a new problem:

$$A^* = \arg\min_A \mathcal{J}^*, \text{ where } \mathcal{J}^* = \mathcal{J} - \lambda \left(\sum_{p=1}^{k}\sum_{q=1}^{k} a_{p,q} - 1\right) \ .$$

By calculating the derivative of $\mathcal{J}^*$ of each $a_{r,s}$, we obtain the formula of the local optimum:

$$\frac{\partial \mathcal{J}^*}{\partial a_{r,s}} = \frac{\partial}{\partial a_{r,s}}\left(\mathcal{J} - \lambda\left(\sum_{p=1}^{k}\sum_{q=1}^{k} a_{p,q} - 1\right)\right) = \frac{\partial \mathcal{J}}{\partial a_{r,s}} - \lambda \frac{\partial}{\partial a_{r,s}}\left(\sum_{p=1}^{k}\sum_{q=1}^{k} a_{p,q} - 1\right) = \frac{\partial \mathcal{J}}{\partial a_{r,s}} - \lambda \ . \quad (13)$$

All three terms of the objective function are dependent on $a_{r,s}$, therefore:

$$\frac{\partial \mathcal{J}}{\partial a_{r,s}} = \lambda_1 \frac{\partial T_1}{\partial a_{r,s}} + \lambda_2 \frac{\partial T_2}{\partial a_{r,s}} + \lambda_3 \frac{\partial T_3}{\partial a_{r,s}} \quad (14)$$

We calculate the derivatives of $T_1$, $T_2$ and $T_3$ on $a_{r,s}$:

$$\frac{\partial T_1}{\partial a_{r,s}} = \frac{\partial}{\partial a_{r,s}} \left[ \sum_{\mu_p \in \mathcal{M}} \sum_{x_i \in \mathcal{C}_p} ||x_i - \mu_p||_{TA} + \sum_{\mu_p \in \mathcal{M}} \sum_{\substack{x_i \in \mathcal{C}_p \\ q \neq p \\ x_i^\phi = x_k^\phi}} \sum_{x_k \in \mathcal{C}_q}^{x_i^t < x_k^t} \left( \beta * e^{-\frac{1}{2}\left(\frac{||x_i^t - x_k^t||}{\delta}\right)^2} \left(1 - a_{p,q}^2\right) \right) \right] =$$

$$= -2a_{r,s} \sum_{\substack{x_i \in \mathcal{C}_r \\ x_i^\phi = x_k^\phi}} \sum_{x_k \in \mathcal{C}_s}^{x_i^t < x_k^t} \left( \beta * e^{-\frac{1}{2}\left(\frac{||x_i^t - x_k^t||}{\delta}\right)^2} \right) =$$

$$= -2a_{r,s}\, pen(\mathcal{C}_r \xrightarrow{\phi} \mathcal{C}_s), \text{ with the notation } pen(\mathcal{C}_r \xrightarrow{\phi} \mathcal{C}_s) = \sum_{\substack{x_i \in \mathcal{C}_r \\ x_i^\phi = x_k^\phi}} \sum_{x_k \in \mathcal{C}_s}^{x_i^t < x_k^t} \left( \beta * e^{-\frac{1}{2}\left(\frac{||x_i^t - x_k^t||}{\delta}\right)^2} \right) .$$

(15)

$$\frac{\partial T_2}{\partial a_{r,s}} = 2a_{r,s}\, ||\mu_r - \mu_s||_{TA} \; . \quad (16)$$

$$\frac{\partial T_3}{\partial a_{r,s}} = 2a_{r,s}\, inter_\phi^2(\mathcal{C}_p, \mathcal{C}_q) \; . \quad (17)$$

By introducing Equations 15, 16 and 17 into Equations 13 and 14, we obtain:

$$\frac{\partial \mathcal{J}^*}{\partial a_{r,s}} = 2a_{r,s} \left( -\lambda_1\, pen(\mathcal{C}_r \xrightarrow{\phi} \mathcal{C}_s) + \lambda_2 ||\mu_r - \mu_s||_{TA} + \lambda_3\, inter_\phi^2(\mathcal{C}_p, \mathcal{C}_q) \right) - \lambda \quad (18)$$

We note $K_{r,s} = -\lambda_1\, pen(\mathcal{C}_r \xrightarrow{\phi} \mathcal{C}_s) + \lambda_2 ||\mu_r - \mu_s||_{TA} + \lambda_3\, inter_\phi^2(\mathcal{C}_p, \mathcal{C}_q)$, we introduce it into Equation 18 and we calculate the fixed point:

$$\frac{\partial \mathcal{J}^*}{\partial a_{r,s}} = 0 \Leftrightarrow 2a_{r,s}K_{r,s} - \lambda = 0 \Leftrightarrow a_{r,s} = \frac{\lambda}{2K_{r,s}} \; . \quad (19)$$

By introducing the constraint in Equation 12 into the Equation 19, we obtain:

$$\sum_{p=1}^{k} \sum_{q=1}^{k} \frac{\lambda}{2K_{p,q}} = 1 \Leftrightarrow \lambda = \frac{1}{\sum_{p=1}^{k} \sum_{q=1}^{k} \frac{1}{2K_{p,q}}} \; . \quad (20)$$

From Equations 19 and 20, we obtain the adjacency matrix update formulas:

$$a_{r,s} = \frac{1}{2K_{r,s} \sum_{p=1}^{k} \sum_{q=1}^{k} \frac{1}{2K_{p,q}}} = \frac{1}{K_{r,s} \sum_{p=1}^{k} \sum_{q=1}^{k} \frac{1}{K_{p,q}}} \ ,$$

$$\text{with } K_{r,s} = -\lambda_1 \ pen(\mathcal{C}_r \xrightarrow{\phi} \mathcal{C}_s) + \lambda_2 ||\mu_r - \mu_s||_{TA} + \lambda_3 \ inter_\phi^2(\mathcal{C}_p, \mathcal{C}_q) \ ,$$

$$\text{and } pen(\mathcal{C}_r \xrightarrow{\phi} \mathcal{C}_s) = \sum_{\substack{x_i \in \mathcal{C}_r \\ x_i^\phi = x_k^\phi}} \sum_{\substack{x_k \in \mathcal{C}_s \\ x_i^t < x_k^t}} \left( \beta * e^{-\frac{1}{2}\left(\frac{||x_i^t - x_k^t||}{\delta}\right)^2} \right) \ .$$

## 3 RUNTIME FOR CLUSPATH AND EVOLUTIONARY ALGORITHM

This section reports runtimes for ClusPath and evolutionary algorithm for determining the ClusPath's parameters. All reported runtimes are influenced by the used platform, as well as software and implementation decisions. Consequently, they should be treated comparatively.

All experiments were run on a machine featuring 12 Intel(R) Xeon(R) CPU E5-2430 cores, each running at 2.20GHz. Each core has hyper-threading activated, resulting in 24 logical cores. Each processor disposes of 16MB of cache, while the system has 64GB of RAM space. The system runs Ubuntu Precise 12.04.5, with a 64bit Linux linux kernel version 3.5.0-43.

Both ClusPath and the evolutionary algorithm were programmed in the Matlab/Octave environment and they were run in Octave 3.6.1, using the OpenBLAS 0.1alpha2.2-3 implementation of BLAS (Basic Linear Algebra Subprograms). For the execution of the evolutionary algorithm, OpenBLAS was forced into single threaded execution. The ClusPath algorithm itself is programmed single-threaded (even if it can be easily parallelized, just like most clustering algorithms). Individuals in an evolutionary population are computed in parallel, using the Octave `multicore` package. Therefore, at each generation, the maximum number of ClusPath executions running in parallel at any given time was 24.

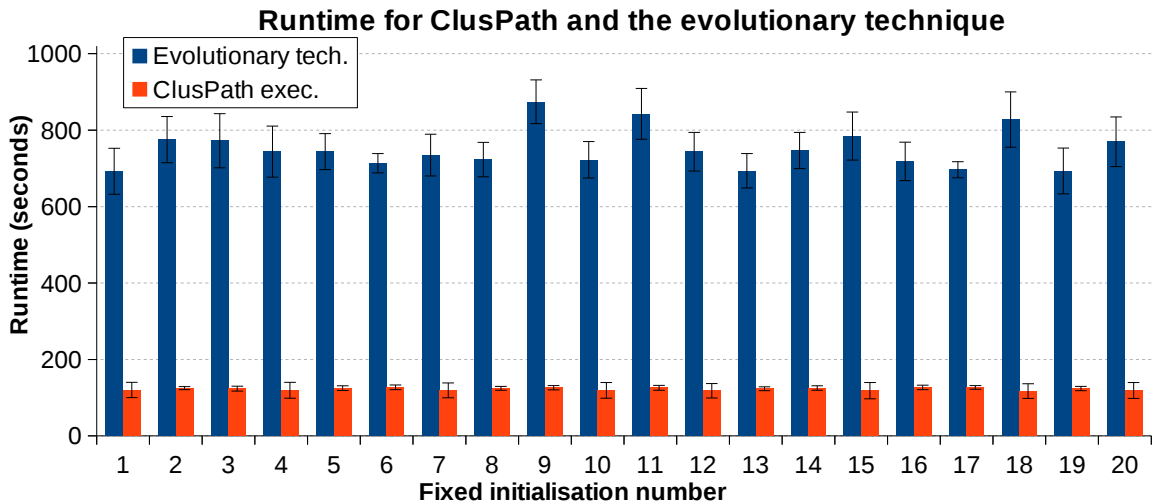

**Supplementary Figure 1.** Runtimes, in seconds, for ClusPath and the Evolutionary Technique, on `CPDS1`. Each algorithm is run 10 times for each of the 20 initializations presented in Sect. **4.2** of MT.

**Supplementary Table 1.** Runtimes, in seconds, for "single" ClusPath and the Evolutionary Technique, on `CPDS1`. Each algorithm is run 10 times, for each of the 20 initializations presented in Sect. 4.2 of MT. The table presents the mean of the 10 runs and the standard deviation (in *italic*). The **Ratio** shows how many time slower was the Evolutionary technique compared to "single" ClusPath. The bottom two lines present aggregated measurements over the 20 initializations.

| Init. no. | Evolutionary algo. | | "single" ClusPath | | Ratio |
|---|---|---|---|---|---|
| 1 | 692,60 | *60,19* | 120,19 | *19,83* | **5,76** |
| 2 | 775,30 | *60,60* | 125,02 | *3,92* | **6,20** |
| 3 | 772,30 | *70,70* | 123,85 | *6,58* | **6,24** |
| 4 | 744,00 | *66,62* | 119,51 | *20,85* | **6,23** |
| 5 | 743,90 | *47,21* | 125,30 | *6,23* | **5,94** |
| 6 | 713,50 | *25,08* | 127,33 | *5,91* | **5,60** |
| 7 | 735,00 | *54,50* | 119,04 | *19,34* | **6,17** |
| 8 | 723,20 | *44,92* | 124,47 | *5,17* | **5,81** |
| 9 | 874,40 | *57,35* | 126,19 | *5,63* | **6,93** |
| 10 | 722,40 | *47,59* | 119,06 | *20,53* | **6,07** |
| 11 | 842,70 | *66,83* | 125,98 | *6,34* | **6,69** |
| 12 | 743,50 | *50,49* | 118,20 | *18,95* | **6,29** |
| 13 | 693,80 | *45,23* | 123,50 | *5,09* | **5,62** |
| 14 | 747,00 | *47,32* | 125,39 | *5,71* | **5,96** |
| 15 | 784,60 | *62,89* | 118,46 | *21,30* | **6,62** |
| 16 | 718,30 | *50,31* | 127,02 | *5,96* | **5,65** |
| 17 | 696,60 | *20,89* | 126,98 | *4,85* | **5,49** |
| 18 | 827,70 | *72,36* | 117,45 | *18,97* | **7,05** |
| 19 | 693,20 | *59,82* | 124,17 | *5,52* | **5,58** |
| 20 | 769,70 | *65,07* | 118,99 | *20,76* | **6,47** |
| **Average** | **750,69** | *53,80* | **122,81** | *11,37* | **6,12** |
| *StDev* | *51,05* | *13,61* | *3,48* | *7,33* | *0,46* |

Figure 1 presents the runtimes of both "single" ClusPath and the evolutionary technique, on `CPDS1`, for each of the 20 initializations presented in Sect. 4.2 of MT. For each initialization, each algorithm is ran 10 times and the means and standard deviation are presented. Furthermore, when running ClusPath alone (*i.e.*, not as an individual in the evolutionary technique), its parameters are chosen randomly from their domain of definition. The purpose of this choice is not to bias the execution of the "single" ClusPath. Fir the first generation of the evolutionary technique, the parameters of the individual runs of ClusPath are chosen similarly, at random from their domain of definition. Therefore, the runtime of ClusPath presented hereafter is comparable to the execution time required for each individual in the evolutionary technique. Table 1 further details these runtimes, presenting also the ratio between the evolutionary technique and "single" ClusPath. More precisely, the evolutionary technique is, in average, $6.12$ times slower than ClusPath. That is consistent with our setup: given that each evolutionary population contains 100 executions of ClusPath and 24 executions are done in parallel, the first generations is roughly equivalent to 4 subsequent executions of ClusPath. Given the elitist technique, each later generation requires a total of ClusPath executions less than 24. Given that the entire optimization requires, in average, 3 generations, we obtain a ratio between the evolutionary technique and ClusPath of approximately 6 times.



\end{document}